\documentclass[a4paper,11pt]{article}
\usepackage{jheppub} 
\usepackage{lineno}

\usepackage{graphicx}
\usepackage{physics}
\usepackage{tikz}
\usepackage{subcaption}
\usepackage{indentfirst}

\newcommand{\half}{\frac{1}{2}}
\newcommand{\quarter}{\frac{1}{4}}
\newcommand{\ve}{\varepsilon}
\newcommand{\expect}[1]{\left\langle {#1} \right\rangle}

\title{\boldmath Perturbative Unorientable JT Gravity and Matrix Models}

\author{Wasif Ahmed and}
\author{Ashton Lowenstein}
\affiliation{Department of Physics and Astronomy, University of Southern California,\\
Los Angeles, CA 90089, U.S.A.}

\emailAdd{wasifahm@usc.edu} \emailAdd{alowenst@usc.edu}

\abstract{We consider an orthogonal polynomial formulation of the double scaling limit of multicritical matrix models in the $\beta=1$ Dyson-Wigner class. They capture the physics of 2D quantum gravity coupled to minimal matter on unorientable surfaces, otherwise called unoriented minimal strings. We derive a formula for the density of states valid to all orders in perturbation theory. We show how to define an interpolation between the multicritical models and that a certain interpolation among an infinite number of them provides an alternative definition of unoriented JT gravity. We discuss the strengths and weaknesses of our formulation.}

\begin{document}
\maketitle
\flushbottom

\section{Introduction}

The double scaling limit of random matrix models can be used to formulate many two dimensional (2D) Euclidean quantum gravity theories. The success of the double scaling limit in describing smooth macroscopic geometries (i.e. gravity) is most easily understood in the context of the 't Hooft large-$N$ limit of matrix/gauge theories \cite{Brezin:1977sv, BREZIN1990144, DOUGLAS1990635, PhysRevLett.64.127, GROSS1990333, HOOFT1974461, KAZAKOV1985295, DAVID198545, BESSIS1980109, Kazakov:1989bc}. In a random matrix model, the ribbon diagram expansion of the partition function is interpreted as literally enumerating tessellations of surfaces of increasingly more complicated topology, with $1/N$ (where $N$ is the size of the matrix) determining the topological expansion parameter. Double scaling the matrix model then amounts to sending $N \to \infty$ while simultaneously shrinking the characteristic area $a$ of a face in a given tessellation. The result is a sum over smooth closed surfaces indexed by their Euler number.

Random matrix theory has a rich history in both physics and math. Its original use by Wigner to describe nuclear physics embodies its statistical nature, where the fundamental object (the matrix) is drawn from a randomly distributed ensemble with probabilities being determined by the matrix potential \cite{Wigner}. While the interpretation in the spirit of 't Hooft connects more directly with geometry and gravity, it appears the Wignerian approach to random matrix theory may shed light on non-perturbative aspects of gravity \cite{Johnson:2022wsr, Johnson:2022hrj}. Nevertheless, the interplay between gravity and random matrix theory also provides a way to expand the math literature on the topic.

A particular 2D theory of interest lately, Jackiw-Teitelboim (JT) gravity, has shed a lot of light on properties of quantum gravity \cite{JACKIW1985343, TEITELBOIM198341, Almheiri:2014cka}. Classical solutions of JT gravity on a closed manifold describe rigid hyperbolic spaces. It is more interesting to consider the theory on a manifold with at least one boundary, in which case the boundary dynamics are described by the Schwarzian theory \cite{Almheiri:2014cka}. In \cite{Saad:2019lba} Saad, Shenker, and Stanford showed that perturbation theory in JT gravity is captured by the double scaling limit of a certain matrix model, which can be cast in terms of a limit of minimal string models \cite{Mertens_2021}. It was subsequently shown in \cite{Johnson:2020heh} that the matrix model describing JT gravity perturbation theory is a special interpolation between an infinite number of multicritical matrix models. Those models have reliable non-perturbative completions, making it possible to fully define JT gravity (and supergravity) in a non-perturbative way \cite{Johnson:2019eik, Johnson:2020exp, Johnson:2021tnl}.

Generalizations of JT gravity (including orientability, defects, supersymmetry, etc.) have been studied from both a field theory \cite{Chamseddine:1991fg, Forste:2017kwy, Fan:2021wsb} and random matrix perspective \cite{Stanford:2019vob, Johnson:2021owr,Johnson:2020heh}. Considering unoriented spacetimes in Euclidean quantum gravity is interesting for several reasons. First, an unoriented bulk spacetime is appropriate for describing a boundary theory with time reversal symmetry in the context of the AdS/CFT correspondence. Second, although the Gaussian Orthogonal Ensemble (GOE) and Gaussian Symplectic Ensemble (GSE) have been studied extensively in the math literature (see for example \cite{forrester2010log}), the double scaling limits of the more general Wigner-Dyson $\beta$-ensembles is not as investigated. Other descriptions of unoriented gravity in two dimensions can be found in \cite{JaumeGomis_2004, OrenBergman_2004, HARRIS1991685}. Since these models describe unoriented minimal strings any effort in this direction provides nice contact between math and physics.

In \cite{Stanford:2019vob} Stanford and Witten discuss unoriented JT gravity and its connection to random matrix theory. They develop perturbation theory using the loop equation formalism, which focuses on the resolvent operator of the random matrix. By applying techniques developed in \cite{Saad:2019lba} they conclude that the presence of even a single cross-cap can cause divergences in otherwise well-behaved objects. See \cite{Stanford:2023dtm} for recent progress relating to the loop equation formulation of unoriented JT gravity. An effort is made in this work to define perturbation theory from a different perspective, in the hopes that a more matrix model-focused approach may resolve the apparent issues with the loop equation formalism. This alternate point of view is met with mixed results, as it turns out to be difficult to extract results from the formalism even for the simplest non-trivial minimal model.

This paper details two methods by which one can analytically compute the large-$N$ and double scaled eigenvalue densities for $\beta = 1$ ensembles with arbitrary polynomial matrix potential. Results for the Gaussian case are known already (see e.g. \cite{mehta2004random, forrester2010log}) and progress has been made in the direction of the string equations for more general models \cite{Brezin:1990dk, Carlisle_2005}. 

The first method we discuss employs intrinsic properties of the aforementioned orthogonal polynomials to compute the $n$-point correlators of the eigenvalues. We argue that the result for the GOE found in \cite{Mahoux} is nearly valid for arbitrary matrix potential, missing only surface term corrections. This formula for the density directly involves the orthogonal polynomials, so the pre-existing double scaling technology can be applied with the inclusion of some extra ingredients. In particular, in \cite{Brezin:1990dk} the authors showed how to double scale the derivative with respect to the matrix eigenvalue $\lambda$
    \begin{equation}
        \frac{d}{d\lambda} \xrightarrow{\text{double scale}} P_k.
    \end{equation}
The operator $P_k$ is the Lax pair of a Sturm-Liouville operator $\mathcal{H} = -\partial^2 + u$, familiar in the KP/KdV hierarchy; in particular it generates the $k^\text{th}$ KdV flow of the operator $\mathcal{H}$. By imposing an auxiliary condition called the string equation one finds that $P_k = \left(\mathcal{H}^{k-\half}\right)_+$, making $P_k$ a $(2k-1)$-order differential operator with polynomials in $u$ and its derivatives as coefficients.

The second method we discuss utilizes a new set of polynomials. This set of functions has a sort of orthogonality relation with respect to a skew-symmetric bilinear form called a ``skew inner product.'' For this reason the functions are called skew orthogonal polynomials. In \cite{Adler_2000} the authors derived an expression for the finite-$N$ eigenvalue density of a $\beta = 1$ matrix model in terms of these polynomials. By adapting results from \cite{Brezin:1990dk} we present the double scaling limit of the density (eqn. \eqref{eqn:density}). The result involves factorizing $P_k$ in terms of a new set of functions $g_i$, making contact with the KdV hierarchy's relationship to more general Drinfeld-Sokolov hierarchies \cite{Drinfeld:1984qv}
    \begin{equation}
        P_k = -\prod_{i = 1}^k \left(\hbar \partial - \half g_i \right)\hbar \partial \prod_{j = 1}^k \left(\hbar\partial + \half g_{k - j}\right).
    \end{equation}
The functions $g_i$ are used to define differential operators $T_k$ and $S_k$ that facilitate a transformation between the skew orthogonal polynomials and a set of associated orthogonal polynomials. Equations of motion for the $g_i$ are determined in terms of $u$ by comparing the factorization to $\left(\mathcal{H}^{k-\half}\right)_+$.

In analogy to the $\beta = 2$ models, a method for implementing an interpolation between minimal models is proposed. The Lax operator $P$ of a such an interpolation is a simple linear combination of the individual Lax operators. The functions $g_i$ do not add linearly in the interpolation, but instead have to be determined by factorizing the entire Lax operator $P$. Utilizing this, we obtain a formal expression for the eigenvalue density of unoriented JT gravity (eqn. \eqref{eqn:UJTdensity}), interpreted as an interpolation between an infinite number of these unoriented minimal models.

The rest of this paper is organized as follows. In {\bf section \ref{scn:matrixmodelsandorthogonalpolynomials}} we review the conventional orthogonal polynomial approach to matrix models. In {\bf section \ref{scn:thebeta1wignerdysonensembles}} we review known results in the finite-$N$ study of $\beta = 1$ Wigner-Dyson models and discuss generalizations of the GOE results to more general matrix potentials. In {\bf section \ref{scn:newresultsinmatrixmodels}} we implement the double scaling limit and obtain a formula for the eigenvalue density, valid to all orders in perturbation theory. The equations of motion for the new functions $g_i$, called new string equations, are discussed. In {\bf section \ref{scn:pureunorientedgravity}} we perform some example calculations in the first few multicritical models to demonstrate the use of our results. In particular we show how to perturbatively determine the functions $g_i$ in terms of the function $u$ for individual multicritical models as well as massive interpolations between them. We then use the WKB approximation to compute perturbative contributions to the double scaled density. A peculiarity involving the GOE is investigated along these lines. Finally, we apply consider the simplest nontrivial unoriented minimal string, which provides a way to test the interpolation formulae. In {\bf section \ref{scn:JTgravity}} we apply our results to obtain a formluation of unoriented JT gravity as an interpolation between an infinite number of minimal models. A formal expression for the eigenvalue density is provided.


\section{Matrix Models and Orthogonal Polynomials} \label{scn:matrixmodelsandorthogonalpolynomials}
We begin by reviewing some basics about random matrix theory (RMT) and establishing some conventions. For a review in the context of two-dimensional gravity see \cite{ginsparg1993lectures}, and for more mathematical reviews \cite{mehta2004random, forrester2010log}.

Given a set of $N \times N$ matrices $\{M\}$, its corresponding random matrix probability density is determined by a function $V$ called the potential and is given by $e^{-N\tr V(M)}dM$, where $dM$ refers to the flat measure on the space of $N^2$ variables. We will take $V$ to be an even polynomial of rank $2k$, and always assume that $N$ is even. The elements of $M$ are random variables, distributed according to this probability density.

An object of central importance in our endeavor is the matrix integral $Z$ given by
    \begin{equation}
        Z = \int dM\,e^{-N\tr V(M)},
    \end{equation}
with respect to which scalar functions of the random matrix, called observables, can be computed as expectation values. It is common for the set of matrices in consideration to have a symmetry, for example being hermitian or real symmetric. In these cases there is a sort of gauge redundancy in $Z$ that can be dealt with by diagonalizing $M$. Performing the change of variables $M \to U^\dagger \operatorname{diag}(\lambda_0, \cdots, \lambda_{N-1}) U$, with $U$ in the appropriate symmetry group, we can rewrite the partition function after integrating out $U$
    \begin{equation}
        Z = C \int \prod_{i = 0}^{N-1} d\lambda_i\,|\Delta(\lambda)|^\beta e^{-N\sum_{i = 0}^{N-1} V(\lambda_i)}, \label{eqn:matrixint}
    \end{equation}
where $\Delta(\lambda) = \prod_{i < j}(\lambda_j - \lambda_i)$ is the Vandermonde determinant. The numerical prefactor $C$ is related to the volume of the symmetry group and so we will drop it. The power $\beta$ is determined by the symmetry group used to diagonalize $M$: for the group $U(N)$ $\beta = 2$ and for $O(N)$ $\beta = 1$. 

The domain of integration for the eigenvalues is also affected by the value of $\beta$. Since $N$ is finite, we can assume that the eigenvalues are ordered. The matrix integral can be defined with a domain of integration that respects the ordering in order to not over count, and to eliminate the absolute value around $\Delta$. However, for even values of $\beta$ we can extend the integration to all of $\mathbb{R}^N$ by dividing by a combinatorial factor to take into account relabelling the indices anytime $\lambda_j - \lambda_i$ changes sign. Having each eigenvalue take any real value is convenient, as will be explored shortly. We are not so fortunate when considering odd values of $\beta$ though, and must use alternate methods.

 It can be shown that $\Delta$ is the determinant of a matrix consisting of powers of the eigenvalues, $\Delta(\lambda) = \det \lambda_i^j$. By taking linear combinations of the rows with real coefficients, we can introduce a family of polynomials $p_i(x)$ so that $\Delta(\lambda) = \det p_j(\lambda_i)$. The polynomials will be normalized so that $p_i(\lambda) = \lambda^i + \cdots.$ We stress here that the introduction of these polynomials is possible for any value of $\beta$, and solely depends on the presence of the Vandermonde determinant.

A convenient choice for the polynomials $p_i$ is to make them orthogonal with respect to a power of the measure $e^{-V}$, for example
	\begin{equation}
		\int_{-\infty}^\infty d\lambda e^{-V(\lambda)} p_i(\lambda)p_j(\lambda) = h_i \delta_{ij}.
	\end{equation}
This is particularly helpful when the values of $\lambda$ can be expanded to all of $\mathbb{R}$. Note that the family of these polynomials is infinite, which will facilitate the large-$N$ limit to be taken later. Define the oscillator wavefunctions
	\begin{equation}
		\psi_i(\lambda) = h_i^{-1/2} e^{-V(\lambda)/2} \, p_i(\lambda). \label{eqn:oscwavefncts}
	\end{equation}
The choice of name has a two-fold motivation. In the Gaussian matrix model, the orthogonal polynomials are the Hermite polynomials and the functions $\psi_i$ are identical to the normalized wavefunctions of the harmonic oscillator. Additionally, we have the Fermi gas, or log-gas, perspective that arises from interpreting the Vandermonde determinant as a logarithmic interaction potential. The eigenvalues behave like fermions and in a Fock space representation each one is represented by a sort of oscillator.

While formally the index $i$ on $\psi_i$ is the same as the index on $\lambda_i$, we know that even for a finite number of eigenvalues there are infinitely many wavefunctions. So, consider a many-body system comprised of $N$ identical fermions. The wavefunction of a single fermion is represented by $\psi_i$, where the index $i$ indicates the energy level occupied by that particular particle. Since the fermions are identical only a single particle can occupy the $i^{\text{th}}$ energy level. By virtue of the fact that the Vandermonde determinant in the matrix model partition function can be written in terms of the orthogonal polynomials, so too can it be written in terms of the first $N$ oscillator wavefunctions. Hence the matrix model takes into account the first $N$ energy levels of the fermions. In a fermionic many-body system, the energy at which all lower energy levels are occupied is called the Fermi level, and the underlying energy levels the Fermi sea. As such, the index value $i = N-1$ is called the Fermi level\footnote{Note that since we begin counting at $i = 0$, the $N$th level is at $i = N-1$.}, and all lower values the Fermi sea.

The oscillator wavefunctions provide an orthonormal basis for the Hilbert space $L^2(\mathbb{R})$ with the standard inner product. Denote the basis as $\{\ket{\psi_i}\}_{i =0}^\infty$. Operators on $L^2$ can thus be decomposed in terms of these functions in the usual way
    \begin{equation}
        \mathcal{O}_{ij} \equiv \mel{\psi_j}{\mathcal{O}}{\psi_i} = \int_{-\infty}^\infty dx\,\psi_j(x) \hat{\mathcal{O}}\psi_i(x).
    \end{equation}
By $\hat{\mathcal{O}}$ we denote the action of the operator on the actual function $\psi_i(x)$.

The polynomials $p_i$ satisfy the following recursion relation
	\begin{equation}
		\lambda p_i(\lambda) = p_{i+1}(\lambda) + R_i p_{i-1}(\lambda),
	\end{equation}
with $R_i = h_i/h_{i-1}$. The corresponding recursion relation for the oscillator wavefunctions is 
	\begin{equation}
		\lambda \psi_i(\lambda) = \sqrt{R_{i+1}}\psi_{i+1}(\lambda) + \sqrt{R_i}\psi_{i-1}(\lambda).
	\end{equation}
The self-adjoint operator $\Lambda$ that implements this transformation on $L^2$ has matrix elements
	\begin{equation}
		\Lambda_{ij} = \sqrt{R_{i+1}} \delta_{j,i+1} + \sqrt{R_{i}} \delta_{j,i-1}.
	\end{equation}

A second operator of interest implements the derivative, which we denote by $L$ in the finite-$N$ regime. One can show that the matrix elements $L_{ij}$ are nonzero in general only when $|i - j| \leq 2k -1$, are determined by the recursion coefficients $R_i$, and are anti-symmetric because $L$ is anti-self-adjoint. The operators $\Lambda$ and $L$ satisfy the canonical commutation relation $[\Lambda,L] \propto 1$.

Since $V$ is an even function it can be shown that each polynomial $p_i$ has definite parity, given by $p_i(-\lambda) = (-1)^i p_i(\lambda)$. This induces the $\mathbb{Z}_2$-grading  $L^2 = \text{span}\{\ket{\psi_{2\alpha}}\} \oplus \text{span}\{\ket{\psi_{2\beta + 1}}\}$. Thus we can conveniently decompose our operators into $2 \times 2$ block form. For example, both $\Lambda$ and $L$ are odd under the grading, so we designate
	\begin{equation}
    \begin{aligned}
        \Lambda &= \left(\begin{matrix} 0 & \ell \\ \ell & 0 \end{matrix} \right), \\
		L &= \left(\begin{matrix} 0 & -c^\dagger \\ c & 0 \end{matrix}\right).
	\end{aligned}
	\end{equation}
Objects that are labelled $0, \dots, N -1$ will carry a Latin index $i,j$ (but note that $k$ is reserved for the order of the matrix potential). When considering $\mathbb{Z}_2$-grading we will decompose some of these objects into even and odd parts, and it will be convenient to write $i = 2\alpha$ or $j = 2\gamma + 1$, respectively. Objects that have definite $\mathbb{Z}_2$ parity will thus carry a Greek index $\alpha,\gamma$ (but $\beta$ is reserved to denote the type of matrix ensemble).

For example, take the operator $L$. By definition the operator $c$ has matrix elements given by $c_{\gamma\alpha} = -L_{2\alpha,2\gamma + 1}$. Since the matrix representing $L$ has $2k - 1$ non-zero off-diagonals, the matrix representing $c$ will have $2k - 1$ non-zero off-diagonals (and the main diagonal will be nonzero as well, which is not true for $L$). The action of the derivative on the even wavefunctions $\psi_{2\alpha}$ can be written
    \begin{equation}
        \psi_{2\alpha}' = -\sum_{\gamma = 0}^{\frac{N}{2} + k - 1} c_{\gamma\alpha} \psi_{2\gamma + 1}. \label{eqn:oscderivative}
    \end{equation}
For this range of $\gamma$ in the sum, some of the matrix elements $c_{\alpha \gamma}$ will be zero depending on the value of $\alpha$, but these bounds of summation are guaranteed to produce all necessary non-zero terms.

The double scaling limit \cite{BREZIN1990144, DOUGLAS1990635, GROSS1990333, KAZAKOV1985295} is a combination of taking the size of the matrix $N \to \infty$ while also focusing the objects of the theory to neighborhoods around specific values. In this limit the small parameter $1/N$ has a renormalized form $\hbar$. The index $i$ is replaced by a continuous variable $x \in (-\infty,\mu]$, where $\mu$ is the Fermi level. The large-$N$ limit confines the spectrum to a single interval\footnote{This is the one-cut case.}, and in the double scaling limit we zoom in on one of the edges of the ineterval, replacing $\lambda$ with a new variable $E \in [0,\infty)$. The oscillator wavefunctions become functions of $\psi(x,E)$, the recursion coefficients $R_i$ are replaced by a function $u(x)$, and the operator $\Lambda$ that implements the recursion becomes a Schrodinger Hamiltonian $\mathcal{H} = -\hbar^2 \partial_x^2 + u(x)$. All together, the recursion relation becomes the Schrodinger equation
    \begin{equation}
        \mathcal{H}\psi(x,E) = E \psi(x,E).
    \end{equation}
The derivative operator $L = \frac{d}{d\lambda}$ is replaced by a differential operator with respect to $x$, called $P_k$. In the Lax formalism of integrable systems, the operators $(\mathcal{H},P_k)$ form the Lax pair of the $k^\text{th}$ model in the KdV hierarchy, where $P_k$ is determined by requiring that the commutator takes a special value\footnote{To be more specific, the commutator is required to be the KdV flow $\frac{\partial u}{\partial t_k}$ with respect to the $k^\text{th}$ KdV time. This detail is outside the scope of this work and will not be used again.}. The unique differential operator satisfying that condition is given by $P_k \propto \left(\mathcal{H}^{k-\half}\right)_+$, where the $+$ subscript denotes keeping only non-negative powers of the derivative $\partial_x$. See Appendix \ref{apdx:PDO} for more details about pseudo-differential operators.

If one is interested in solving the Schrodinger equation, the function $u$ must be determined first. In Hermitian matrix models, the equation of motion is precisely the canonical commutation relation $[P_k,\mathcal{H}] = 1$, which is the double scaled version of the relation satisfied by $\Lambda$ and $L$ mentioned previously. The outcome is expressed in terms of the Gelfand-Dikii differential polynomials $R_k$. The equation of motion, called a string equation for historical reasons, is
    \begin{equation}
        R_k + x = 0.
    \end{equation}
Two important examples of Gelfand-Dikii polynomials that will be used later are
    \begin{equation}
        R_1 = u,\quad R_2 = -\frac{\hbar^2}{3}u'' + u^2. \label{eqn:GDpolys}
    \end{equation}
More general models can be defined by introducing coupling constants $t_k$ and considering the equation of motion
    \begin{equation}
        \sum_k t_k R_k + x = 0.
    \end{equation}
Such a model is called an interpolation between the individual multicritical models. This can be obtained from the canonical commutation relation by defining the generalized operator $P = \sum_k t_k P_k$.


\section{The \texorpdfstring{$\beta=1$}{b} Wigner-Dyson ensembles} \label{scn:thebeta1wignerdysonensembles}

To construct an enumeration of unoriented surfaces using a matrix integral, we restrict ourselves to real symmetric matrices $H$, which falls under the $\beta = 1$ $O(N)$ symmetry class. The matrix $H$ is in the bi-fundamental representation $\mathbf{N} \otimes \mathbf{N}$ of $O(N)$. The two fundamental representations are identical, making the indices of $H$ indistinguishable. As such, it will be more convenient not to keep track of upper and lower indices. 

To motivate considering this theory as unoriented, we will briefly touch on the 't Hooft diagrams of the theory. The Gaussian action can be cast in propagator language
    \begin{equation}
        N\tr(H^2) = \frac{N}{2} \sum_{i,j,m,n = 0}^{N-1} H_{im}\Big( \delta_{in}\delta_{mj} + \delta_{ij}\delta_{mn}\Big)H_{jn}.
    \end{equation}
This implies that the propagator is $\frac{2}{N}(\delta_{in}\delta_{mj} + \delta_{ij}\delta_{mn})$ for $\beta = 1$ ensembles. The second term comes from the fact that the indices are indistinguishable for our matrices: $H_{ij} = H_{ji}$.

When one adds higher order terms to the potential and performs a diagrammatic expansion of $Z$, the presence of the new term in the propagator adds twists to the diagrams, which can only be drawn on unoriented surfaces\footnote{This is, of course, up to equivalences between even numbers of crosscaps and handles.}. In the large-$N$ limit, observables in the theory will have a topological $1/N$ expansion. For instance, the free energy defined by $Z = e^{\tilde{F}}$ can be expanded as
    \begin{equation}
        \tilde{F} = \sum_{g = 0}^\infty \sum_{c = 0}^\infty N^{2-2g-c} \tilde{F}_{g,c}.
    \end{equation}
The number $\chi(g,c) = 2-2g-c$ is the Euler characteristic of the diagram, corresponding to a tessellation of a surface with $g$ handles and $c$ crosscaps. In the double scaling limit, the quantity\footnote{We remove the tilde to distinguish between the two scaling regimes.} $F_{g,c}$ represents the contribution to the free energy of that topology.

Since $\beta = 1$ is odd, we must take care to specify a domain of integration in $Z$ that allows us to remove the absolute value on $\Delta$. We write
	\begin{equation}
		Z = \int_{\mathcal{I}}  \prod_{i = 0}^{N-1} d\lambda_i \Delta(\lambda) e^{-N\sum_{j = 0}^{N-1} V(\lambda_j)}, \label{eqn:beta1matrixint}
	\end{equation}
where $\mathcal{I}$ denotes the ordering $-\infty < \lambda_{N -1}\leq \lambda_{N-2} \leq \cdots \leq \lambda_1 \leq \lambda_0 < \infty$. General $n$-point functions of the eigenvalues can be computed using an idea pioneered by Dyson \cite{doi:10.1063/1.1703773} and applied to the GOE by Mehta and Mahoux \cite{Mahoux}, called integration over alternate variables. The analysis begins by introducing two arbitrary functions $u$ and $v$, and considering the following expectation value with respect to the matrix ensemble
	\begin{equation}
		G_N(u,v) = \expect{\prod_{\alpha = 0}^{\frac{N}{2} - 1} u(\lambda_{2\alpha}) \prod_{\beta = 0}^{\frac{N}{2} - 1} v(\lambda_{2\beta + 1}) }.
	\end{equation}
Further progress will be made by judiciously introducing a set of polynomials whose properties simplify the calculation. There are two natural choices one can make, which are the subjects of the following subsections.


\subsection{Orthogonal Polynomial Approach}

First, introduce a set of orthogonal polynomials $p_i(\lambda)$ satisfying
    \begin{equation}
        \int_{-\infty}^\infty d\lambda\,e^{-2NV(\lambda)}p_i(\lambda)p_j(\lambda) = h_i \delta_{ij}.
    \end{equation}
Notice that the integration measure includes a different power of $e^{-NV}$ than used previously. Unless stated otherwise, the polynomials $p_i$ will be orthogonal with respect to this new measure. Consequently, the oscillator wavefunctions will be defined as
    \begin{equation}
        \psi_i(\lambda) = \frac{1}{\sqrt{h_i}}p_i(\lambda) e^{-NV(\lambda)}.
    \end{equation}

Introduce the functions
	\begin{equation}
		F_i(v;\lambda_{2\alpha}) = \int_{-\infty}^{\lambda_{2\alpha}} d\lambda_{2\alpha + 1} v(\lambda_{2\alpha + 1})\psi_i(\lambda_{2\alpha + 1}).
	\end{equation}
 By absorbing the factors of $v$ into the determinant and taking linear combinations of the columns of the matrix, it can be shown that $G_N$ is given by
	\begin{equation}
	\begin{aligned}
		&G_N(u,v) = C_N  \int_{-\infty}^\infty d\lambda_0 e^{-NV(\lambda_0)} u(\lambda_0) \int_{-\infty}^{\lambda_0}d\lambda_2 e^{-NV(\lambda_2)} u(\lambda_2)\cdots \\
  &\cdots\int_{-\infty}^{\lambda_{N - 4}}d\lambda_{N-2}e^{-NV(\lambda_{N-2})}u(\lambda_{N-2}) \det(m(v;\lambda_{2\alpha})),
	\end{aligned}
	\end{equation}
where $C_N$ is a numerical factor depending on the normalizations $h_i$ and the matrix $m$ is given by
	\begin{equation}
	\begin{aligned}
		m_{i,2\alpha} &= p_i(\lambda_{2\alpha}), \\
		m_{i,2\alpha + 1} &= F_i(v;\lambda_{2\alpha}).
	\end{aligned}
	\end{equation}
The integrand is now symmetric under the swap of any two variables (due to the invariance of the determinant under swapping an even number of rows and columns), so we can enlarge the domain of integration to $(-\infty, \infty)$ for each eigenvalue at the cost of a numerical prefactor, which we disregard. Subsequently, the integrals over each eigenvalue are decoupled and $G_N$ can be written
	\begin{equation}
		G_{N}(u,v) = C'_N \sum_{\sigma \in S_N} \prod_{\alpha = 0}^{\frac{N}{2}-1}J_{\sigma(2\alpha),\sigma(2\alpha + 1)}(u,v),
	\end{equation}
where
	\begin{equation}
		J_{ij}(u,v) = \int_{-\infty}^\infty dx \int_{-\infty}^\infty dy\,u(x)v(y)\psi_{[i}(x)\psi_{j]}(y),
	\end{equation}
and the brackets denote anti-symmetrization. This determines the correlator $G_N$ as the Pfaffian of $J$.

By taking linear combinations of the rows and columns of $J$, we can put it in a $2 \times 2$ block form. Begin by setting $J_{2\alpha,2\gamma}(u,v) = f_{\alpha\gamma}(u,v)$. Next, define
	\begin{equation}
	\begin{aligned}
		g_{\alpha\gamma}(u,v) &= -\sum_{\sigma = 0}^{\frac{N}{2}+k-1} c_{\sigma\gamma}J_{2\alpha,2\sigma+1}(u,v) \\
			&= \int_{-\infty}^\infty dx \int_{-\infty}^\infty dy\,u(x)v(y)\Big(\psi_{2\alpha}(x)\psi'_{2\gamma}(y) - \psi_{2\alpha}(y)\psi_{2\gamma}'(x)\Big).
	\end{aligned}
	\end{equation}
Notice that the values of the dummy index $\sigma$ go outside of the ``$N \times N$ block'' in index space in which the matrix model technically lives. The impact of this on the eigenvalue density will be discussed shortly. Finally define
	\begin{equation}
	\begin{aligned}
		\mu_{\alpha\gamma}(u,v) &= \sum_{\mu,\nu = 0}^{\frac{N}{2}+k-1} c_{\mu\alpha}J_{2\mu+1,2\nu+1}c_{\nu\gamma} \\
			&= \int_{-\infty}^\infty dx \int_{-\infty}^\infty dy\,u(x)v(y)\Big(\psi'_{2\alpha}(x)\psi'_{2\gamma}(y) \\ 
             &\quad - \psi'_{2\alpha}(y)\psi'_{2\gamma}(x)\Big).
	\end{aligned}
	\end{equation}
After some rearranging to ensure antisymmetry, we find the Pfaffian can be written
	\begin{equation}
		\text{Pf}(J) \to \text{det}^{1/2}\left(\begin{matrix} f_{\alpha\gamma} & g_{\alpha\gamma} \\ -g_{\gamma\alpha} & \mu_{\alpha\gamma} \end{matrix} \right),
	\end{equation}
and hence the correlator $G_N$ can be expressed in terms of this new determinant. If $u,v$ are both even functions, the diagonal blocks of this new matrix are zero and the Pfaffian reduces to $\det(g)$.

We now use the two functions $u,v$ as sources and treat $G_N$ as a generating function for the eigenvalue correlators
	\begin{equation}
		\rho_n(x_0,\dots,x_{n-1}) =\frac{N!}{(N-n)!} \int_{-\infty}^\infty d\lambda_{n} 
  \int_{-\infty}^\infty d\lambda_{N-1}\,\prod_{i < j} |\lambda_i - \lambda_j|e^{-N\sum_{k = 0}^{N-1}V(\lambda_k)}.
	\end{equation}
These correlators are related to $G_N$ via \cite{mehta2004random}
	\begin{equation}
		G_N(1+a,1+a) = \sum_{n = 1}^{N} \frac{1}{n!}\int_{-\infty}^\infty \prod_{i = 0}^{n-1}a(x_i)dx_i 
            \rho_n(x_0,\dots,x_{n-1}) 
	\end{equation}
In particular, this means that to compute the eigenvalue density $\rho_1(\lambda) \equiv \tilde{\rho}(\lambda)$, we take one functional derivative of $G_N$ and set the source to $0$. Mehta shows that for the GOE
	\begin{equation}
		\tilde{\rho}(\lambda) = \sum_{\alpha = 0}^{\frac{N}{2}-1} \left[ \psi_{2\alpha}(\lambda)^2 - \psi'_{2\alpha}(\lambda) \int_0^\lambda d\lambda' \psi_{2\alpha}(\lambda')\right].\label{eqn:finitedensity}
	\end{equation}

The result \eqref{eqn:finitedensity} nearly holds for $k > 1$. Denoting the density for the $k$th multicritical model by $\tilde{\rho}^{(k)}$, the result is really
    \begin{equation}
        \tilde{\rho}^{(k)} = \tilde{\rho}^{(1)} + \text{ surface terms}, \label{eqn:generalfinitedensity}
    \end{equation}
where the oscillator functions used in $\tilde{\rho}$ are $k$-dependent as well. The additional terms, called surface terms here, are localized around $\alpha = N/2$ and arise from the fact that we included terms outside of the Fermi sea to obtain $\psi_{2\alpha}'$ when rewriting the matrix $J$. The number of extra terms is set by $k$ in the upper limit of summation in \eqref{eqn:oscderivative}. The surface terms, denoted by $\mathcal{K}(\lambda)$, have the form
    \begin{equation}
        \mathcal{K}(\lambda) = \sum_{\alpha = 0}^{\frac{N}{2}-1} \mathcal{N}_\alpha \sum_{\sigma = \frac{N}{2}}^{\frac{N}{2}-1+k} c_{\sigma \alpha} \psi_{2\sigma + 1}(\lambda),\quad \quad \mathcal{N}_\alpha = \int_{-\infty}^\infty \psi_{2\alpha}(\lambda')d\lambda'.
    \end{equation}
In the simplest case $k = 2$,
    \begin{equation}
        \mathcal{K}(\lambda) = \sum_{\alpha = 0}^{\frac{N}{2}-1} \mathcal{N}_\alpha \Big(c_{\frac{N}{2},\alpha} \psi_{N+1}(\lambda) + c_{\frac{N}{2}+1,\alpha} \psi_{N+3}(\lambda)\Big).
    \end{equation}
It is not possible to determine any of the quantities in this formula in closed form for finite $N$.

\subsection{Skew Orthogonal Polynomial Approach}

While the orthogonal polynomial approach bears some fruit when applied to the $\beta = 1$ models, there is another set of functions more naturally suited to solving the problem. We begin again with (\ref{eqn:beta1matrixint}) and compute the correlator $G_N(u,u)$, setting $u = v$ for convenience. We still arrive at the expression
    \begin{equation}
	\begin{aligned}
		G_N(u,v) = C_N & \int_{-\infty}^\infty d\lambda_0 e^{-NV(\lambda_0)} u(\lambda_0) \int_{-\infty}^{\lambda_0}d\lambda_2 e^{-NV(\lambda_2)}u(\lambda_2)\cdots \\
		& \cdots \int_{-\infty}^{\lambda_{N - 4}}d\lambda_{N-2}e^{-NV(\lambda_{N-2})}u(\lambda_{N-2}) \det(m(v;\lambda_{2\alpha})),
	\end{aligned}
    \end{equation}
where now we define
    \begin{equation}
	\begin{aligned}
		m_{i,2\alpha} &= q_i(\lambda_{2\alpha}), \\
		m_{i,2\alpha + 1} &= \tilde{F}_i(u;\lambda_{2\alpha})= \int_{-\infty}^{\lambda_{2\alpha}} d\lambda' e^{-NV(\lambda)}q_i(\lambda').
	\end{aligned}
    \end{equation}
The polynomials $q_i(\lambda)$ are monic and chosen to obey the following relation \cite{Brezin:1990dk, Adler_2000, Mahoux}
    \begin{equation}
        \langle q_{2\alpha},q_{2\gamma + 1}\rangle_s \equiv \iint_{-\infty}^\infty d\lambda d\lambda'\,e^{-NV(\lambda) - NV(\lambda')} \ve(\lambda - \lambda')q_i(\lambda)q_j(\lambda') = - r_\alpha \delta_{\alpha\gamma}, 
    \end{equation}
where $\ve(\lambda - \lambda') = \text{sgn}(\lambda - \lambda')$, and with all other $\langle q_i,q_j \rangle_s = 0$. The bilinear form $\langle \cdot,\cdot\rangle_s$ is skew-symmetric, and hence is referred to as a skew inner product. The polynomials $q_i$ are called skew orthogonal polynomials. The correlator $G_N(u,u)$ is once again given by the pfaffian of a matrix $\tilde{J}$ given by
    \begin{equation}
        \tilde{J}_{ij} = \langle uq_i,uq_j \rangle_s.
    \end{equation}
If we set $u = 1$, this computes the full matrix integral.

In analogy to the oscillator wavefunctions $\psi_i$ defined above, introduce the skew oscillator wavefunctions
    \begin{equation}
        \zeta_i(\lambda) = r_i^{-1/2} e^{-NV(\lambda)}q_i(\lambda).
    \end{equation}
Notice that the matrix $\tilde{J}$ has the $2\times2$ block structure
    \begin{equation}
        \tilde{J}[u] = \left(\begin{matrix} f_{\alpha \gamma}[u] & g_{\alpha \gamma}[u] \\ - g_{\gamma \alpha}[u] & \mu_{\alpha \gamma}[u] \end{matrix} \right),
    \end{equation}
where
    \begin{equation}
        \begin{aligned}
        f_{\alpha \gamma}[u] &\equiv \tilde{J}_{2\alpha,2\gamma}[u], \\
        g_{\alpha \gamma}[u] &\equiv \tilde{J}_{2\alpha,2\gamma + 1}[u], \\
        \mu_{\alpha \gamma}[u] &\equiv \tilde{J}_{2\alpha + 1,2\gamma + 1}[u].
        \end{aligned}
    \end{equation}
The functions $f,g$ and $\mu$ referenced here are not the same as the functions with the same names used in the orthogonal polynomial approach. Consider the function $u = 1 + a$ where $a$ is small. Then the $2 \times 2$ structure of $\tilde{J}$ can be written (schematically) as
    \begin{equation}
        \tilde{J} = \left(\begin{matrix} \epsilon f & 1 + \epsilon \nu \\ -(1 + \epsilon \nu) & \epsilon \mu \end{matrix}\right),
    \end{equation}
with $\epsilon$ small and related to the function $a$. The pfaffian pf$\tilde{J}$ admits a straightforward expansion in $\epsilon$ (see Appendix A.7 in \cite{mehta2004random}). The density can once again be extracted from $G_N(1 + a)$ by taking a functional derivative with respect to $a$. The result is \cite{Adler_2000, Mahoux}
    \begin{equation}
        \tilde{\rho}(\lambda) = \sum_{\alpha = 0}^{\frac{N}{2} - 1} \Big(\zeta_{2\alpha}(\lambda)[\hat{\ve}\cdot\zeta_{2\alpha + 1}](\lambda) - \zeta_{2\alpha+1}(\lambda)[\hat{\ve}\cdot\zeta_{2\alpha}](\lambda)\Big). \label{eqn:densityskewwavefunctions}
    \end{equation}
This expression for the eigenvalue density is exact for any value of $N$. This is in contrast to \eqref{eqn:finitedensity}, which required the inclusion of $k$-dependent surface terms to be correct.

The construction of skew orthogonal polynomials and their properties are explored in detail in \cite{Adler_2000}. However, the ultimate goal of this work is to reach the double scaling limit, and there does not appear to be a process through which one can directly do this to the skew orthogonal polynomials. This difficulty is circumvented by once again appealing to orthogonal polynomials. 

The orthogonal polynomials $p_i$ introduced above form a basis for the ring of polynomials. The skew orthogonal polynomials $q_i$ can be expanded in terms of the $p_i$: there exists a lower triangular matrix $O$ with 1's on the diagonal such that
    \begin{equation}
        q_i = \sum_{j = 0}^i O_{ij}p_j. \label{eqn:ptoq}
    \end{equation}
Representing the skew oscillator wavefunctions with the kets $\ket{\zeta_i}$, the defining relation for the polynomials $q_i$ is then given by
    \begin{equation*}
        \langle q_i, q_j \rangle_s = 2\sqrt{r_ir_j}\mel{\zeta_i}{\hat{\ve}}{\zeta_j},
    \end{equation*}
where $\hat{\ve}$ is the integral operator with kernel $\half \ve(\lambda - \lambda')$ and the $L^2$ inner product is used. An implication of this is that the skew inner product can be implemented by an operator acting on $L^2$ with the measure $e^{-2NV}$. By taking into account the normalizations, the skew oscillator wavefunctions can be expressed in terms of the oscillator wavefunctions in a way related to \eqref{eqn:ptoq}
    \begin{equation}
        \zeta_i = \sum_{j = 0}^i \tilde{O}_{ij}\psi_j,\quad \tilde{O}_{ij} = \sqrt{h_j/r_i}O_{ij}.
    \end{equation}
The skew oscillator wavefunctions have the same $\mathbb{Z}_2$-grading as the oscillator wavefunctions, so the matrix $\tilde{O}$ must be even under the $\mathbb{Z}_2$-grading. Define the matrices $a,b$ so that
    \begin{equation*}
        \tilde{O} = \left( \begin{matrix} a & 0 \\ 0 & b \end{matrix}\right).
    \end{equation*}
In \cite{Brezin:1990dk} they show that $a,b$ are related to the derivative matrix $c$ by
    \begin{equation}
        c = b^Ta. \label{eqn:cbarelation}
    \end{equation}
Hence the eigenvalue density is written (schematically) as
    \begin{equation}
        \tilde{\rho} \sim (a\cdot \psi_{\text{even}})\cdot (\hat{\ve}\cdot b\cdot \psi_{\text{odd}}) - (b\cdot \psi_{\text{odd}}) \cdot (\hat{\ve}\cdot b\cdot \psi_{\text{even}}). \label{eqn:densityschematic}
    \end{equation}

To conclude, it should be noted that $\tilde{\rho}$ is the diagonal part of a larger function \cite{Adler_2000}. Define
    \begin{equation}
        s(\lambda,\lambda') = \sum_{\alpha = 0}^{\frac{N}{2} - 1} \Big(\zeta_{2\alpha}(\lambda)[\hat{\ve}\cdot\zeta_{2\alpha + 1}](\lambda') - \zeta_{2\alpha+1}(\lambda')[\hat{\ve}\cdot\zeta_{2\alpha}](\lambda)\Big).    
    \end{equation}
The analog of the self-reproducing kernel familiar from studies of $\beta = 2$ theories is given by
    \begin{equation}
        f_1(\lambda,\lambda') = \left(\begin{matrix} s(\lambda,\lambda') & I(\lambda,\lambda') \\
        \partial_\lambda s(\lambda,\lambda') & s(\lambda,\lambda')\end{matrix} \right),
    \end{equation}
where
    \begin{equation}
        I(\lambda,\lambda') = \half \int_{-\infty}^\infty s(\lambda,z)\ve(z-\lambda')dz - \half \ve(\lambda-\lambda').
    \end{equation}
The eigenvalue density is given by
    \begin{equation}
        \tilde{\rho}(\lambda) = \lim_{\lambda' \to \lambda} \text{qdet}f_1(\lambda,\lambda'),
    \end{equation}
where qdet denotes the quaternionic determinant (see \cite{Mahoux}, for example). The result follows from the fact that $I(\lambda,\lambda) = 0$.

\subsection{Previous Double Scaling Results}

The term ``double scaling limit'' used in this paper corresponds to the ``soft edge'' of an eigenvalue distribution in the math literature. The soft edge limits of numerous matrix models with Gaussian potentials, including the $\beta = 1,2,4$ theories, can be determined by using known facts about the Hermite polynomials. See \cite{Forste:2017kwy} for a review. In particular, the double scaled eigenvalue density for the GOE can be expressed in terms of the Airy function
    \begin{equation}
        \rho_{\text{GOE}}(E) = \rho_{\text{Airy}}(E) + \half \text{Ai}(-E) \left(1 - \int_{-\infty}^E \text{Ai}(-E')dE'\right), \label{eqn:GOEdensity}
    \end{equation}
where $\rho_{Airy}$ is the eigenvalue density for the double scaled GUE (also known as the Airy model).

The double scaling limits of the operators $a,b$ introduced above were computed in \cite{Brezin:1990dk}. In the normalization used here, they are
    \begin{equation}
        \begin{aligned}
            a &\to T_k = \hbar^{k-1}\partial^{k-1} + \sum_{i = 0}^{k-2}\hbar^ig_i\partial^i, \\
            b &\to S_k = \hbar^k\partial^k + \sum_{i = 0}^{k-1}\hbar^ih_i \partial^i.
        \end{aligned} \label{eqn:generalTSdef}
    \end{equation}
The double scaled version of \eqref{eqn:cbarelation} is $S_k^\dagger T_k= P_k$. The authors of \cite{Brezin:1990dk} outline an algorithm to determine the coefficients in $T_k,S_k$ which we will describe below. Using this, they compute the free energy of a pure unoriented gravity theory (corresponding to the single minimal model $k = 2$).


\section{New Results in Matrix Models}\label{scn:newresultsinmatrixmodels}

\subsection{Double Scaled Density} 

We begin by double scaling the modified version of the finite-$N$ GOE density given in \eqref{eqn:generalfinitedensity}. First, in the strictly large-$N$ limit we change the index $i$ into a continuous variable $X \in [0,1]$. Then introduce a small parameter $\delta$ that picks out the scaling parts
    \begin{equation}
    \begin{aligned}
        X &\to \delta^4 x ,\quad \lambda \to \delta^2E, \\
        \psi_i(\lambda) &\to \psi(x,E),\quad \frac{d}{d\lambda} \to \delta^{-2}P_k.
    \end{aligned}
    \end{equation}
In particular in the large-$N$ limit, the eigenvalue distribution for these models settles into a finite window $[-\lambda_c,\lambda_c]$ which, due to the symmetry of the theory, is symmetric about the origin. The full scaling ansatz for the eigenvalue is $\lambda = \lambda_c - \delta^2 E$; using this as a coordinate transformation implies that integrals over $\lambda$ have the scaling part
    \begin{equation}
        \int d\lambda \to -\delta^2 \int dE,
    \end{equation}
where the bounds of integration are updated on a case-by-case basis.

By changing the large-$N$ integration over $\alpha/N$ to $i/N$ we introduce a factor of $\half$. The double scaled density is then
    \begin{equation}
        	\rho(E) = \half \int_{-\infty}^0\left[\psi(x,E)^2 - P_k\psi(x,E)\int_E^{\infty}\psi(x,E')dE'\right]dx + \mathcal{F}_k[\psi](0,E). 
    \end{equation}
The double scaling limit of the surface terms from (\ref{eqn:generalfinitedensity}) are denoted by $\mathcal{F}_k[\psi]$. Since before double scaling there are $\mathcal{O}(1)$ surface terms around $\alpha = N/2$, $\mathcal{F}_k[\psi]$ should involve derivatives of $\psi$ evaluated at the Fermi surface $x = 0$. For larger $k$, there are more surface terms before double scaling and hence higher order derivatives in the double scaling limit. 

In order to double scale the $\beta = 1$ density written in terms of the skew oscillator wavefunctions \eqref{eqn:densityskewwavefunctions} we need to discuss the double scaling limit of the operator $\hat{\ve}$. Consider the application of $\hat{\ve}$ to functions with definite parity. The results are
    \begin{equation}
    \begin{aligned}
         \text{even}:& \quad \hat{\ve}\cdot f(\lambda) = \int_0^\lambda f(\lambda') d\lambda', \\
         \text{odd}:& \quad \hat{\ve}\cdot f(\lambda) = \int_\lambda^{\lambda_c} f(\lambda') d\lambda',
    \end{aligned}
    \end{equation}
where we've assumed that $f$ only has support on $[-\lambda_c,\lambda_c]$. By the scaling ansatz for the $\lambda$ integral,
    \begin{equation}
    \begin{aligned}
        \text{even}:& \quad \hat{\ve}\cdot f(\lambda) \xrightarrow{\text{double scale}} \delta^2 \int_E^\infty f(E')dE',\\
        \text{odd}:& \quad \hat{\ve}\cdot f(\lambda) \xrightarrow{\text{double scale}} -\delta^2 \int_0^E f(E')dE'.
    \end{aligned}
    \end{equation}

The formula for the finite-$N$ density in \eqref{eqn:densityskewwavefunctions} involves the skew wavefunctions, which are related to the oscillator wavefunctions via the transformations $a$ and $b$ as depicted in \eqref{eqn:densityschematic}. The operators $a,b$ double scale to differential operators $T_k,S_k$, which are order$-(k-1)$ and $k$ respectively. The double scaled limit of the relationship between $\zeta$ and $\psi$ is expressed as\footnote{The powers of $\delta$ appearing in these relationships may not actually both be $-1$. What is important is that the sum of the two powers is $-2$, as it is for the operator $P_k$.}
    \begin{equation}
    \begin{aligned}
        \text{even}:& \quad \zeta_{2\alpha}(\lambda) = [a\cdot \psi]_{2\alpha}(\lambda) \xrightarrow{\text{double scale}} \delta^{-1} T_k\psi(x,E) \\
        \text{odd}:& \quad \zeta_{2\alpha+1}(\lambda) = [b\cdot \psi]_{2\alpha+1}(\lambda) \xrightarrow{\text{double scale}} \delta^{-1} S_k\psi(x,E) \\
    \end{aligned}
    \end{equation}

Putting together these pieces, we find the expression for the double scaled $\beta = 1$ density
    \begin{equation}\label{eqn:density}
        \rho(E) = \frac{1}{2\hbar} \int_{-\infty}^0dx\Bigg[T_k\psi(x,E)\left(\int_0^E S_k\psi(x,E')dE'\right) + S_k\psi(x,E)\left(\int_E^\infty T_k \psi(x,E')dE'\right)\Bigg].
    \end{equation}
This expression is exact and does not ignore any surface terms. The normalization is chosen to extract the finite piece after double scaling.

The full expression for the density gives us the opportunity to roughly determine what the surface terms were in the orthogonal polynomial approach. Consider the first term in the above formula. If we were to integrate by parts with respect to $x$ in order to shift $T_k$ inside the $E'$ integral, we would pick up a contribution at $x = 0$ and get the term
    \begin{equation}
        \psi(x,E)\int_0^E T^\dagger_k S_k \psi(x,E')dE',
    \end{equation}
where the adjoint is taken in $L^2(\mathbb{R})$. Now, the specific combination $T_k^\dagger S_k$ is proportional to $\hat{P}_k$, which is canonically conjugate to the Hamiltonian $\mathcal{H}$ for which $\psi(x,E)$ is an eigenfunction. Thus $P_k$ has a representation $P_k \sim \partial_E$, and
    \begin{equation}
        \psi(x,E)\int_0^E T^\dagger_k S_k \psi(x,E')dE' \propto \psi(x,E)\Big(\psi(x,E) -  \psi(x,0)\Big),
    \end{equation}
which produces the $\psi^2$ term familiar from the study of $\beta = 2$ theories.

Consider the second term in the density. Once again we integrate by parts to shift the location of $\hat{T}_k$. This picks up more terms at $x = 0$ and gives
    \begin{equation}
        T_k^\dagger S_k \psi(x,E)\int_E^\infty \psi(x,E') dE' \propto P_k\psi(x,E)\int_E^\infty \psi(x,E') dE'.
    \end{equation}
Being more careful about signs, we conclude
    \begin{equation}
        \rho(E) = \half \int_{-\infty}^0dx\,\Bigg[\psi(x,E)^2 - P_k\psi(x,E)\int_E^\infty \psi(x,E')dE'\Bigg] + \text{ surface terms},
    \end{equation}
where the surface terms are generated by moving the operator $T_k$ around. In the case $k = 2$,
    \begin{equation}
    \begin{aligned}
    -\frac{1}{4\hbar E}&\left(P_k\psi(0,E)\psi(0,0) - \psi(0,E)P_k\psi(0,0)\right)\\ & + \frac{1}{2\sqrt{2}\hbar} \left[ \psi(0,E)\int_0^E S_2 \psi(0,E')dE' + S_2\psi(0,E)\int_E^\infty \psi(0,E')dE' \right].
    \end{aligned}
    \end{equation}

\begin{figure}
\centering
    \begin{subfigure}[t]{0.48\textwidth}
          \centering
          \includegraphics[width=\textwidth]{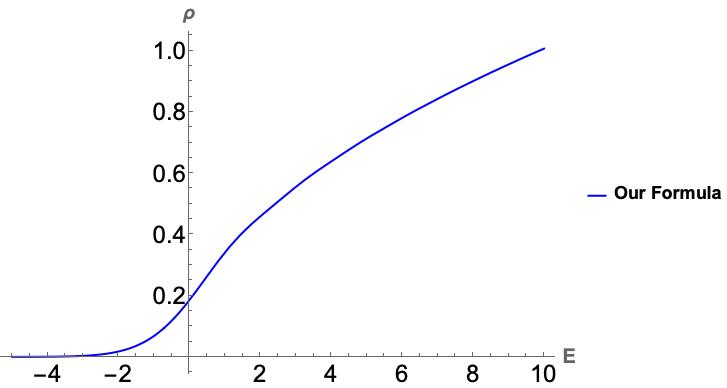}
          \caption{Plot of \eqref{eqn:dslGOEdensity}.}
          \label{fig:dslGOEdensity}
    \end{subfigure}
    \begin{subfigure}[t]{0.48\textwidth}
          \centering
          \includegraphics[width=\textwidth]{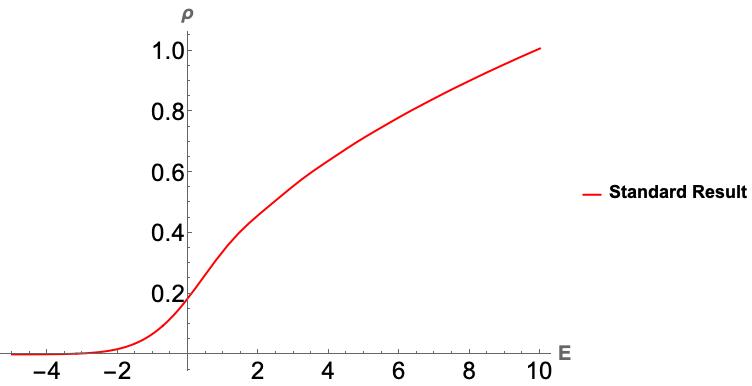}
          \caption{Plot of \eqref{eqn:GOEdensity}.}
          \label{fig:GOEdensisty}
    \end{subfigure}
    \begin{subfigure}[t]{0.48\textwidth}
          \centering
          \includegraphics[width=\textwidth]{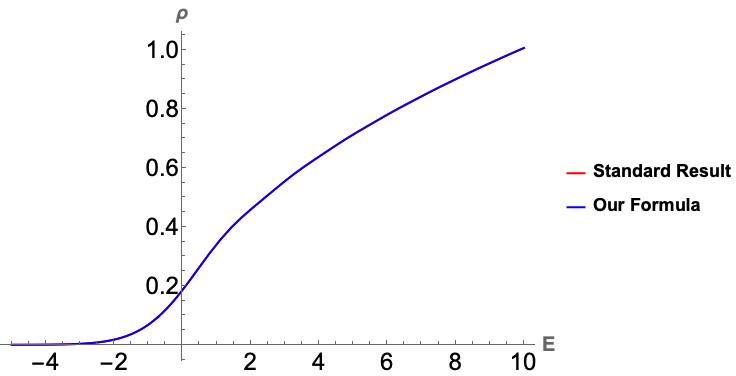}
          \caption{Both the plots.}
          \label{fig:bothplots}
    \end{subfigure}
\caption{Our result from \eqref{eqn:dslGOEdensity} exactly overlaps with the standard result \eqref{eqn:GOEdensity}, as can be seen from $(c)$. Two plots are indistinguishable.}
\label{fig:GOEplots}
\end{figure}

One thing that is obscured by this analysis is what happens in the GOE. For that theory, where $k = 1$, the operator $T_1$ is a constant, meaning it costs nothing to move it around inside the integral. Thus there are no surface terms in that case
    \begin{equation}
        \rho_{\text{GOE}}(E) = \half \int_{-\infty}^0dx\,\Bigg[\psi(x,E)^2 - P_1\psi(x,E)\int_E^\infty \psi(x,E')dE'\Bigg]. \label{eqn:dslGOEdensity}
    \end{equation}
 We can confirm numerically that \eqref{eqn:dslGOEdensity} matches the previously known result given in eqn. \eqref{eqn:GOEdensity} and the plots are shown in \eqref{fig:GOEplots}.

\subsection{New String Equations}

It is evident from \eqref{eqn:density} that we have to determine the operators $T_k,S_k$, and there is a well defined algorithm to determine them outlined in \cite{Brezin:1990dk}. For the $k^{\text{th}}$ minimal model we can write the following factorization equation\footnote{Technically we will be working with the operator $-iP_k$, if $P_k = \left(\mathcal{H}^{k - \half}\right)_+$. This factor of $i$ can be restored by modifying the factorization in terms of $T_k$ and $S_k$, but the functions $g_i$ would remain unchanged, and so would all of our other results. The details can be found in Appendix \ref{apdx:PDO}.}
\begin{equation}
    P_k =-\prod_{i = 1}^{k-1}\left(\hbar\partial - \half g_i\right)\hbar\partial \prod_{j = 1}^{k-1}\left(\hbar\partial + \half g_{k-j}\right)\label{eqn:factorization}.
\end{equation}
The operator $P_k$ is $(2k-1)$-order differential operator where the coefficients are functions of $u$ and its derivatives. Notice how this differs from writing $P_k = S_k^\dagger T_k$ with $T_k,S_k$ previously defined in eqn. (\ref{eqn:generalTSdef}). There, the two individual operators were defined in terms of separate families of functions. which we called $g_i,h_i$. By imposing that the factorization be anti-self-adjoint and setting some integration constants to 0 (see \cite{Brezin:1990dk}) one can reduce the number of functions so that $T_k,S_k$ depend on the same family of functions. The new factorization in eqn. (\ref{eqn:factorization}) amounts to setting
    \begin{equation}
    \begin{aligned}
        S_k &= \prod_{j = 1}^{k-1} \left(\hbar^2 \partial^2 + \frac{\hbar}{2}g_{k-j}\partial + \frac{\hbar}{2}g_{k-j}'\right),\\
        T_k &= \prod_{j = 1}^{k-1}\left(\hbar \partial + \half g_j\right). \label{eqn:TandSdefs}
    \end{aligned}
    \end{equation}
Expanding both sides of \eqref{eqn:factorization} and equating powers of $\partial$ we get $(2k-1)$ coupled differential equations for $(k-1)$ $g_i$ functions. These equations will not all be independent, and we can find $(k-1)$ independent differential equations.

While we do not have exact forms for each equation obtained using the algorithm outlined above for arbitrary $k$, it is straightforward to show that the equation determined by $\partial^{2k-3}$ is always\footnote{This exact form is technically only true for a single multicritical model. See the following discussion of massive interpolations for more.} \cite{Brezin:1990dk} (see Appendix \ref{apdx:PDO} for details)
    \begin{equation}
        \left(k - \half\right)u = \sum_{j = 1}^{k-1}\left[-\hbar jg_j' + \frac{1}{4}g_j^2\right]. \label{eqn:geqn1}
    \end{equation}
Moreover the equation determined by $\partial^{2k-4}$ is simply the derivative of the above equation. This implies that $g_0 \sim \sqrt{u_0}$, even for more complicated massive interpolations.

The preceding algorithm can be adapted to consider an interpolation up to the $k^{\text{th}}$ model with coupling constants $t_i$. Since the order of $P_i$ as a differential operator increases with $i$, the total order of the superposition $P = \sum_i t_i P_i$ as a differential operator is set by $k$. Since \eqref{eqn:geqn1} is determined by $\partial^{2k-3}$, the highest order term with a nontrivial coefficient, the only two operators that can contribute to that equation are $P_{k}$ and $P_{k-1}$. Therefore the proper way to modify (\ref{eqn:geqn1}) is
    \begin{equation}
         \left(k - \half \right)t_{k}u + t_{k-1} = t_{k}\sum_{j = 1}^{k-1}\left[-\hbar jg_j' + \frac{1}{4}g_j^2\right].
    \end{equation}
The right-hand side is unchanged because the total differential operator being factorized has order $2k-1$. The other independent differential equations are determined by looking at the coefficients of lower order derivatives. The next one comes from $\partial^{2k-5}$, and therefore will explicitly include the coupling constants $t_k, t_{k-1}$, and $t_{k-2}$.


\section{Topological and Pure Unoriented Gravity} \label{scn:pureunorientedgravity}

The simplest minimal model, $k = 1$, is dual to topological gravity. Although it is not a genuine theory of surfaces it is interesting to study in its own right, and provides the opportunity to do many analytical calculations. The first nontrivial minimal model, $k = 2$, is dual to pure gravity. 

\subsection{The GOE}

We can use the fact that the wavefunctions $\psi(x,E)$ and the eigenvalue density $\rho$ are both known exactly in terms of the Airy function to test some of the framework above. Note that the GOE does not require an extra function $g$ to compute the density. 

The Schrodinger equation for the $k = 1$ model is\footnote{Strictly speaking this is true for the GUE. The Schrodinger equation for the GOE is related to this by a rescaling of the energy. See the appendix for discussion.}
    \begin{equation}
        \hbar^2 \partial_x^2 \psi + (x + E)\psi = 0.
    \end{equation}
Taking the standard WKB form of the wavefunction, we find the two possibilities
    \begin{equation}
        \psi(x,E) = \frac{1}{\sqrt{2\pi \hbar}} (x + E)^{-1/4}\left[ 1 - \left( \frac{i5\hbar}{48(x + E)^{3/2}} + \frac{5\hbar^2}{64 (x + E)^3} \right) \right] e^{\pm i \frac{2}{3\hbar} (x + E)^{3/2}} + \cdots. \label{eqn:WKBwavefunction}
    \end{equation}
Of course, the differential equation can be solved exactly in terms of the Airy function, whose Taylor series expansion in small-$\hbar$ is a linear combination of the two perturbative solutions obtained above
    \begin{equation}
    \begin{aligned}
        \hbar^{-2/3}\text{Ai}\left(-\hbar^{-2/3}(x + E)\right) &\approx \frac{1}{\sqrt{\pi\hbar}}(x + E)^{-1/4}e^{-\frac{5\hbar^2}{64}(x + E)^{-3}}\\
        &\times \cos\left[ \frac{2}{3\hbar}(x + E)^{3/2} - \frac{5\hbar}{48}(x + E)^{-3/2} - \frac{\pi}{4} \right].
    \end{aligned}
    \end{equation}
The WKB expansion is inherently an expansion in small $\hbar$. However, the non-perturbative oscillating piece has to be dealt with since it does not simplify under the assumption that $\hbar \ll 1$. We circumvent that with the following trick. The formula for the density is ambiguous as it does not allow for the possibility that the wavefunction $\psi$ is complex. The formula's derivation from the matrix model does not fix this ambiguity, so it may be done ad hoc. The simplest option is to replace one wavefunction in each term with its conjugate. Taking the permutations of which wavefunctions get replaced ensures the outcome is real. Upon doing this, the non-perturbative pieces generally cancel out.

For the Gaussian theory,
    \begin{equation}
        \begin{aligned}
            T \propto 1, \\
           S \propto \frac{\partial}{\partial x}.
        \end{aligned}
    \end{equation}
Using the GOE wavefunctions $\psi(x,2\sqrt{2}E)$ one finds
    \begin{equation}
        \rho(E) = \frac{2^{3/4}\sqrt{E}}{\pi \hbar} + \frac{7\hbar}{2^{3/4}256 \pi E^{5/2}} + \cdots.
    \end{equation}
We note three things about the outcome. First, the disk level density has the normalization one would expect from the saddle point analysis (see appendix). Second, the $\mathcal{O}(\hbar^1)$ term has a different coefficient than the $\beta = 2$ density, but the same functional dependence on $E$. The two are not related to each other by the simple rescaling of $E$ -- this is not an issue because we do not expect the theories to be related in a simple way past the leading order in perturbation theory. Third, the $\mathcal{O}(\hbar^0)$ term has a coefficient of 0 according to the calculation. If this were truly a (perturbative) topological expansion in a theory of unoriented surfaces, then the surface with one boundary and one crosscap (i.e. the Mobius strip) would appear at this order. There is a non-vanishing, non-perturbative contribution to the $\mathcal{O}(\hbar^0)$ integral, but it actually contributes at $\mathcal{O}(\hbar^1)$. As a corollary of this, there are no terms in the perturbative series at $\mathcal{O}(\hbar^{2n})$. As we show in the appendix, the absence of the crosscap term is not unique to the WKB approach, but is also seen in the loop equation formalism.

\subsection{The \texorpdfstring{$k = 1$}{k1} to \texorpdfstring{$k = 2$}{k2} Interpolation}

Take an interpolation between $k = 1$ and $k = 2$ with coupling constants $t_1$ and $t_2$. The momentum operator for this interpolation is given by
    \begin{equation}
        P = t_2\hbar^3 \partial^3 + \left(\frac{3t_2}{2}u + t_1\right)\hbar \partial + \frac{3t_2}{4}\hbar u'.
    \end{equation}
There is only one independent function $g$ in the factorization of $P$, so
    \begin{equation}
    \begin{aligned}
        S_2 &=\hbar^2\partial^2+\frac{1}{2}g\hbar \partial +\frac{1}{2}\hbar g',\\
        T_2 &=t_2 \left(\hbar \partial +\frac{1}{2}g\right).\label{STk2}
    \end{aligned}
    \end{equation}
The independent equation of motion for $g$ is
    \begin{equation}
        \frac{3t_2}{2}u+t_1 = t_2 \left(\hbar g' + \quarter g^2\right).
    \end{equation}
Each function will have a perturbative expansion in $\hbar$\footnote{The subscripts here are used to denote the order of the perturbative expansion of the single function $g$, whereas before the subscript enumerated a family of different functions in the factorization of $P$.}
    \begin{equation}
    \begin{aligned}
        u &= u_0 + \hbar^2 u_2 + \hbar^4 u_4 + \cdots \\
        g &= g_0 + \hbar g_1 + \hbar^2 g_2 + \cdots.
    \end{aligned}
    \end{equation}
The first orders in the expansion of $g$ are given by
    \begin{equation}\label{eqn:gfunctions}
    \begin{aligned}
        g_0 &= \pm \sqrt{\frac{4t_1}{t_2}+6u_0}, \\
        g_1 &= \frac{3t_2u'_0}{2t_1+3t_2u_0},\\
        g_2 &= \frac{3\sqrt{t_2} \left(8 t_1^2 u_2+8 t_1t_2 u_0''+24 t_1t_2 u_0 u_2+12 t_2^2u_0 u_0''-15t_2^2 u_0'^2+18 t_2^2u_0^2 u_2\right)}{2 \sqrt{2}  (2 t_1+3t_2 u_0)^{5/2}}.
    \end{aligned}
    \end{equation}
It is easy to show that as $t_1 \to 0$, the perturbative expansion smoothly transitions to the single $k = 2$ case.

In the following two subsections we will use these results to study the $k = 2$ theory, which is dual to pure unoriented gravity, and the $(2,3)$ unoriented minimal string.

\subsection{Pure Unoriented Gravity}

Consider just the $k = 2$ minimal model. The leading contribution to the density of states for the $k^\text{th}$ minimal model is proportional to $E^{k - \half}$, so we expect to see $E^{3/2}$ behavior. The density admits a topological expansion in powers of $\hbar$,
\begin{equation}
    \rho(E) = \sum_{g,c = 0}^\infty \hbar^{2g+c-1} \rho_{g,c}(E), \label{eqn:densitytopologicalexpansion}
\end{equation}
where $g$ is the number of handles of the surface and $c$ is the number of crosscaps. By convention we define $\rho_{0,0} \equiv \rho_0$.

By keeping the leading order contributions from $T_2,S_2$ we find the disk density

\begin{equation}\label{diskdensity}
    \rho_0(E)=\int_{-\infty}^{0}dx\left[\quarter \partial(g_0 \psi)\int_E^{\infty}(g_0 \psi) dE'-\quarter (g_0 \psi)\int_0^E\partial(g_0\psi)dE'\right].
\end{equation}
Isolating the first subleading contributions gives the crosscap density
    \begin{equation}
    \begin{aligned}
        \rho_{0,1}(E)&=\int_{-\infty}^{0}dx\Bigg[\quarter \partial(g_1\psi)\int_E^{\infty}(g_0\psi)dE'+\quarter \partial(g_0\psi)\int_E^{\infty}(g_1\psi)dE'\\
        &-\quarter (g_1\psi)\int_0^{E}\partial(g_0\psi)dE'-\quarter (g_0\psi)\int_0^{E}\partial(g_1\psi)dE'\\
        &+\half \partial(g_0\psi)\int_E^{\infty}(\partial\psi)dE' + \half \partial^2\psi\int_E^{\infty}(g_0\psi)dE'\\
        &-\half (\partial\psi)\int_0^{E}\partial(g_0\psi)dE'-\half (g_0\psi)\int_0^{E}(\partial^2\psi)dE'\Bigg]. \label{crosscapdensity}
    \end{aligned}
    \end{equation}
The functions $g_0$ and $g_1$ are obtained from \eqref{eqn:gfunctions} by setting $t = 0$.

Both \eqref{diskdensity} and \eqref{crosscapdensity} contain more information than is desired, unless we compute the wavefunctions perturbatively using WKB analysis. The leading order contribution to $u$, found by solving $R_2 + x = 0$ (see \eqref{eqn:GDpolys} for the definition of $R_2$), is $u_0 = \sqrt{-x}$. The WKB analysis yields
    \begin{equation}
        \psi(x,E) = \frac{1}{\sqrt{\pi\hbar}}\left(E - \sqrt{-x}\right)^{-1/4}\cos\left[\frac{4}{15\hbar} \sqrt{E - \sqrt{-x}}\left(2E^2 + E\sqrt{-x} + 3x  \right) \right].
    \end{equation}
Unfortunately it is difficult to maintain neither analytical nor numerical control over the calculation of the density using the WKB wavefunctions. In particular, the combination of the complicated $E$-dependence of the WKB wavefunction and the presence of the $E$ integrals in the formula for the density renders analytical computation very difficult. Ordinarily in a Hermitian matrix model, one would use an averaging argument to get rid of the oscillatory part of the WKB wavefunction to compute perturbative contributions to the density. However in this case the $E$-integrals interact with this oscillatory functions and impact the counting of powers of $\hbar$ in a nontrivial way, as is evident from performing the analogous calculation in the GOE.

As a last resort we appeal to the power of the non-perturbative numerical framework described in \cite{Johnson:2021tnl}, which we summarize here. It has been known since the early 90s that ensembles of positive Hermitian matrices provide a viable nonperturbative completion for models of two-dimensional quantum gravity (see \cite{Dalley:1991vr} for example).
The string equation for $u$ in such a model is modified to be
    \begin{equation}
        u\mathcal{R}^2 - \frac{\hbar^2}{2}\mathcal{R}'' + \frac{\hbar^2}{4}(\mathcal{R}')^2 = 0, \label{eqn:bigstringeq}
    \end{equation}
where $\mathcal{R} = \sum_{k} t_k R_k + x$. Equation  \eqref{eqn:bigstringeq}, with the coupling constants specified, can be solved numerically by putting the system in a box and imposing the boundary conditions
    \begin{equation}
        \mathcal{R}[u(x \to -\infty)] = 0,\quad u(x \to +\infty) = -\frac{1}{4x^2}.
    \end{equation} 
Then the Schrodinger equation for this potential is discretized and solved using the methods of \cite{Numerov}. 

In this work, a spatial grid size of $\Delta x = .03$ is used. The presence of the walls needed to obtain numerical solutions constrains the maximum trusted energy. The average spacing between energy levels is $\Delta E \approx .02$.

\begin{figure}[h!]
    \centering
    \includegraphics[width=10cm, height=6cm]{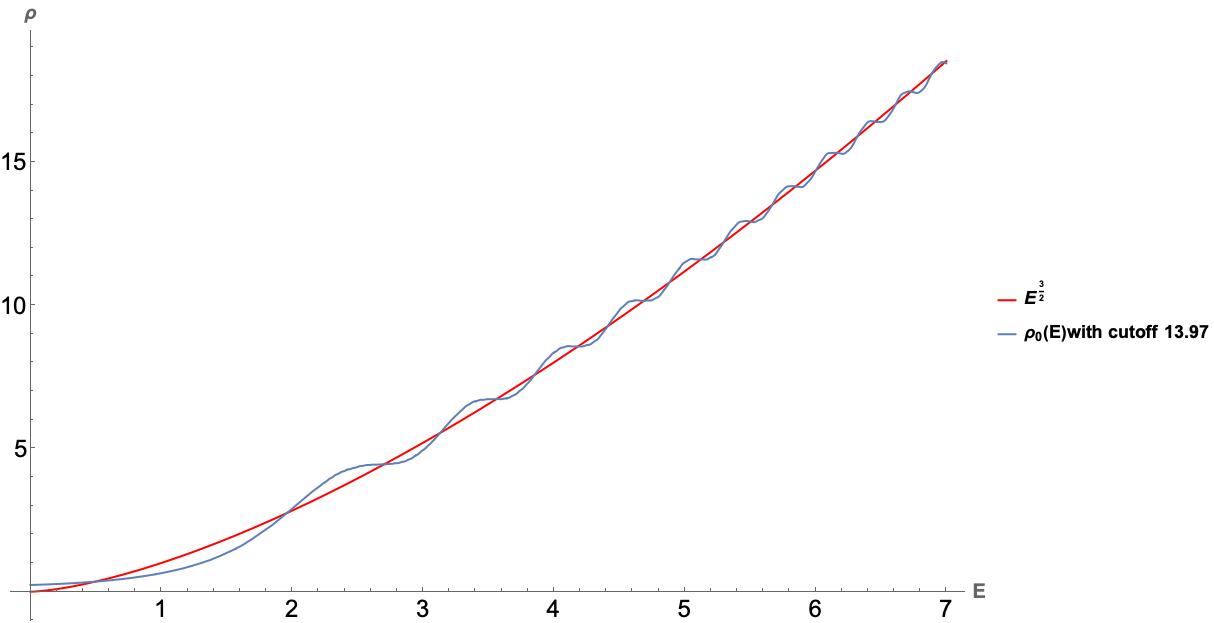}
    \caption{$\rho_0$ vs E curve. We can see that $\rho_0$ perfectly wraps around $E^{\frac{3}{2}}$ in this energy window.}
    \label{fig:disk_density}
\end{figure}

The numerically determined wavefunctions are inserted into eq. \eqref{diskdensity} to produce the disk level density in fig. (\ref{fig:disk_density}).
The curve is oscillating because the wavefunctions are non-perturbative. In a more complete non-perturbative framework the oscillations reveal the discrete nature of the quantum system \cite{Johnson:2022wsr}. However, here they are meaningless because the formalism is only meant to be trusted at the level of perturbation theory. 

\begin{figure}[h!]
    \centering
    \includegraphics[width=10cm, height=6cm]{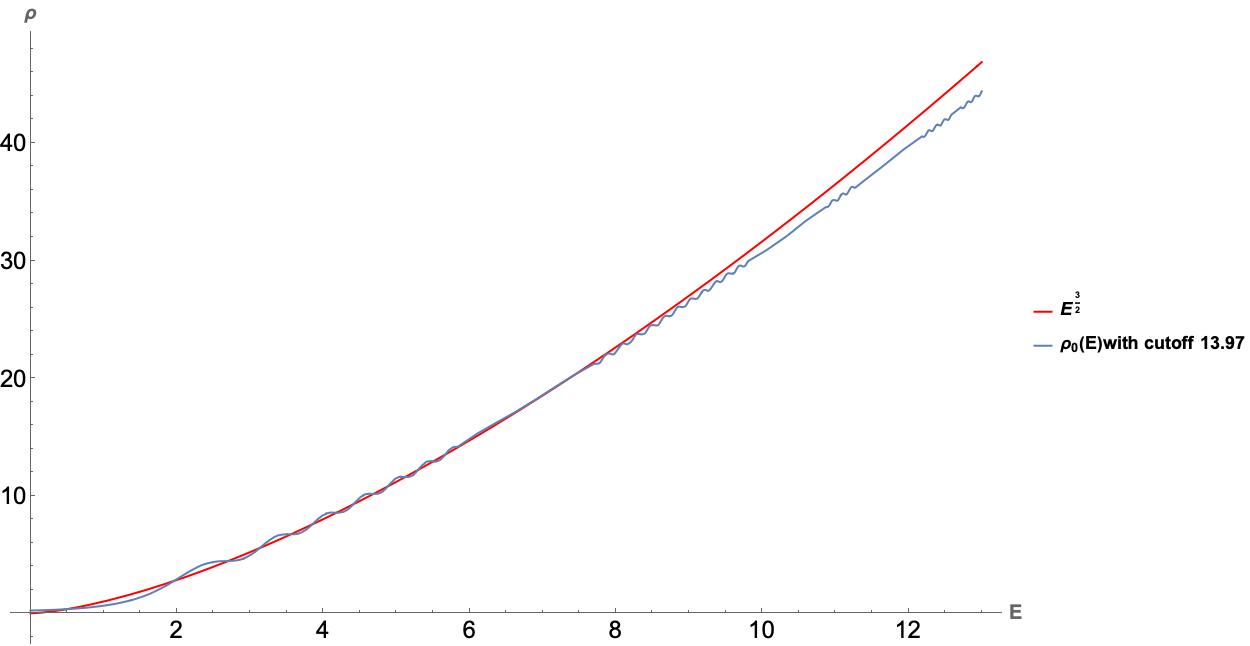}
    \caption{$\rho_0$ vs E curve. We can see the deviation at higher energies.}
    \label{fig:deviation}
\end{figure}

If we plot $\rho_0$ up to higher energies we see a deviation of the curve from $E^{\frac{3}{2}}$ as shown in figure \eqref{fig:deviation}.

\begin{figure}
\centering
    \begin{subfigure}[t]{0.49\textwidth}
          \centering
          \includegraphics[width=\textwidth]{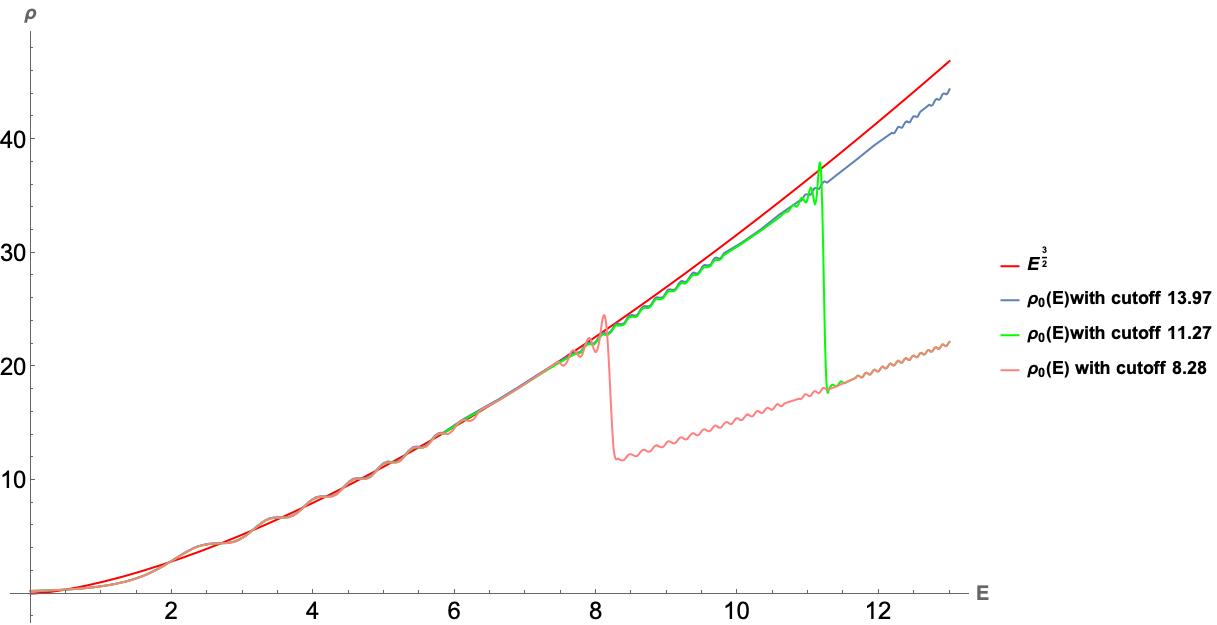}
          \caption{$\rho_0$ with different cutoffs.}
          \label{fig:cutoffs013}
    \end{subfigure}
    \begin{subfigure}[t]{0.49\textwidth}
          \centering
          \includegraphics[width=\textwidth]{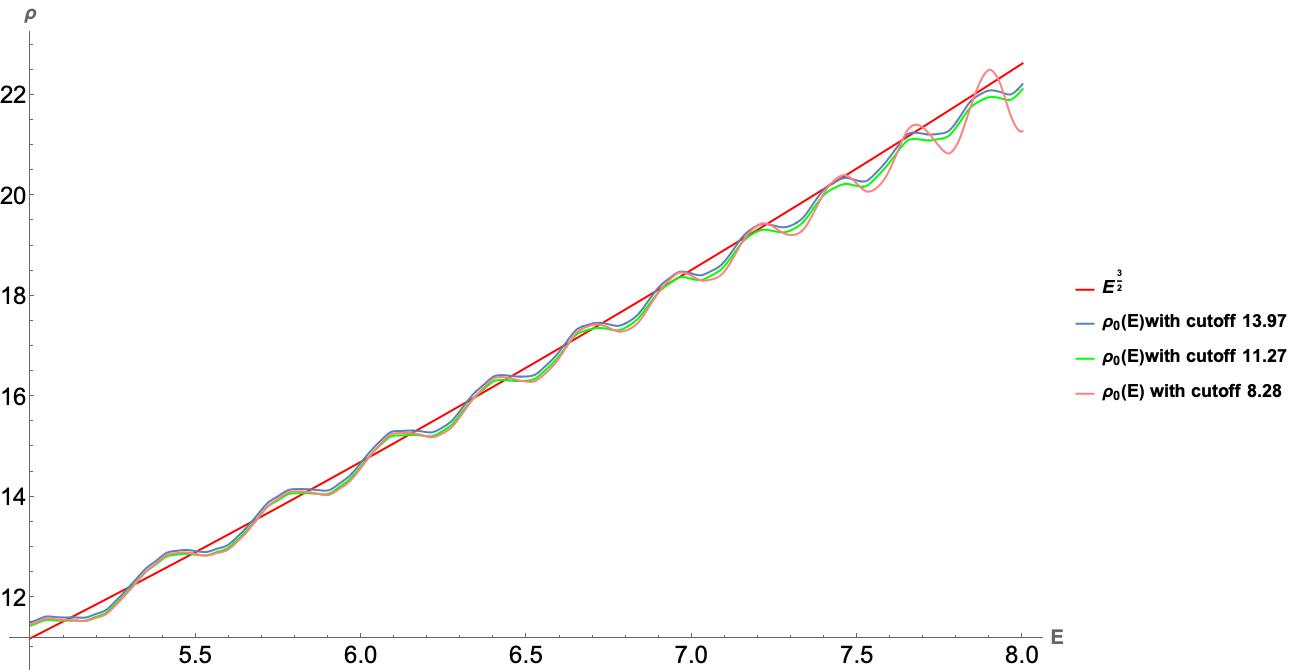}
          \caption{Zoomed in}
          \label{fig:zoomed}
    \end{subfigure}
\caption{For lower energies all the curves with different cutoffs merge together. The curve deviates earlier as the cutoff is decreased. If we zoom in, we can see that for a particular energy the curve with a lower cutoff is more off from the $E^{\frac{3}{2}}$ curve.}
\label{fig:different_cutoffs}
\end{figure}

The deviation is there because we have an integral from $E$ to $\infty$ in \eqref{diskdensity}, and this upper limit was replaced by the highest energy eigenvalue we had in our numerical solution. In this plot, that cutoff was roughly $E_{\text{max}} = 14$. This argument can be solidified by checking the plot of the function with different cutoffs, i.e. upper limits, as shown in figure \eqref{fig:different_cutoffs}.

\begin{figure}[h!]
    \begin{subfigure}[t]{0.49\textwidth}
        \centering
        \includegraphics[width=\textwidth]{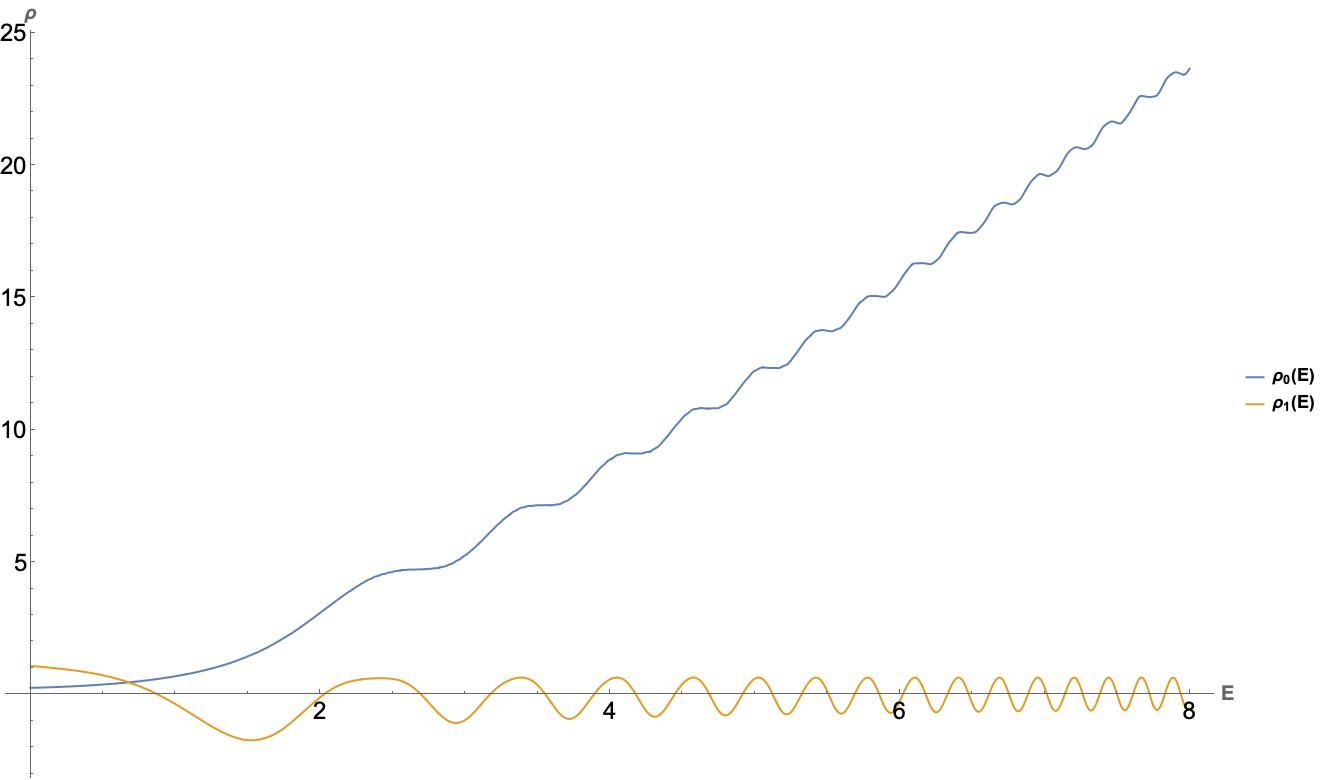}
        \caption{$\rho_0$ and $\rho_{0,1}$ separately}
        \label{fig:crosscap_density}
    \end{subfigure}
    \begin{subfigure}[t]{0.49\textwidth}
        \centering
        \includegraphics[width=\textwidth]{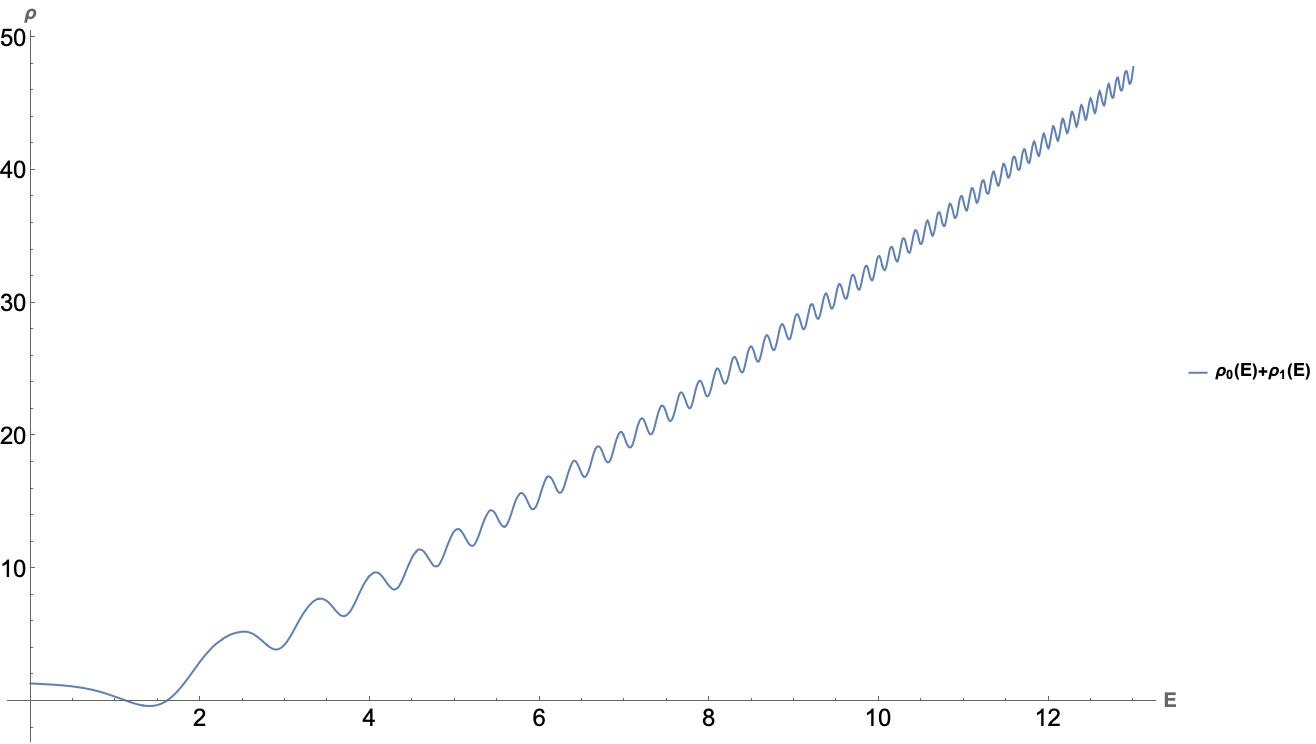}
        \caption{$\rho_0 + \rho_{0,1}$}    
    \end{subfigure}
    \caption{The first unoriented contribution to the eigenvalue density is small compared to the disk level contribution, even with $\hbar = 1$. The combined contribution to the total density still has fictitious oscillations.}\label{fig:total_density}
\end{figure}

We can similarly plot $\rho_{0,1}$ by using the numerically determined wavefunctions in eq. \eqref{crosscapdensity}. To compare the values of $\rho_0$ and $\rho_{0,1}$, we plot them together in figure \eqref{fig:total_density}. We can see that $\rho_0$ is more dominant, and $\rho_1$ is just like a damped oscillation around the $E$ axis, as it should be for a perturbative correction. The sum $\rho_0 + \rho_{0,1}$ still oscillates with the approximate $E^{3/2}$ behavior of the leading contribution.

The amplitude of $\rho_{0,1}$ is decreasing as $E \to \infty$. The topological expansion of the density in \eqref{eqn:densitytopologicalexpansion} is equivalently an expansion in large $E$, so at large energy we expect to recover the disk contribution.

\subsection{The \texorpdfstring{$(2,3)$}{(2,3)} Unoriented Minimal String}

The $(2,2p-1)$ oriented minimal string is described by a matrix model with the coupling constants \cite{Mertens_2021}
    \begin{equation}
        t_k = \frac{\pi^{2k-2}}{k!(k-1)!}\frac{4^{k-1}(p + k - 2)!}{(p-k)!(2p-1)^{2k-2}}.
    \end{equation}
In particular, for the choice $p = 2$ the nonzero couplings are $t_1 = 1$ and $t_2 = 4\pi^2/9$. The leading order contribution to $u$ is found to be
    \begin{equation}
        u_0 = \frac{3 \left(\sqrt{9-16 \pi ^2 x}-3\right)}{8 \pi ^2}.
    \end{equation}
It is important to consider this model for two reasons. First, it represents a proof-of-concept for the interpolation technology described here, that we apply to unoriented JT gravity in the following section, because the leading eigenvalue density can be computed analytically. Second, it is well known that JT gravity is in some sense a $p \to \infty$ limit of the $(2,2p-1)$ minimal string (see \cite{Mertens_2021}, for example). Thus studying the first nontrivial interpolation in this family is a step toward the full theory.

\begin{figure}[h!]
    \centering
    \includegraphics[width=10cm, height=6cm]{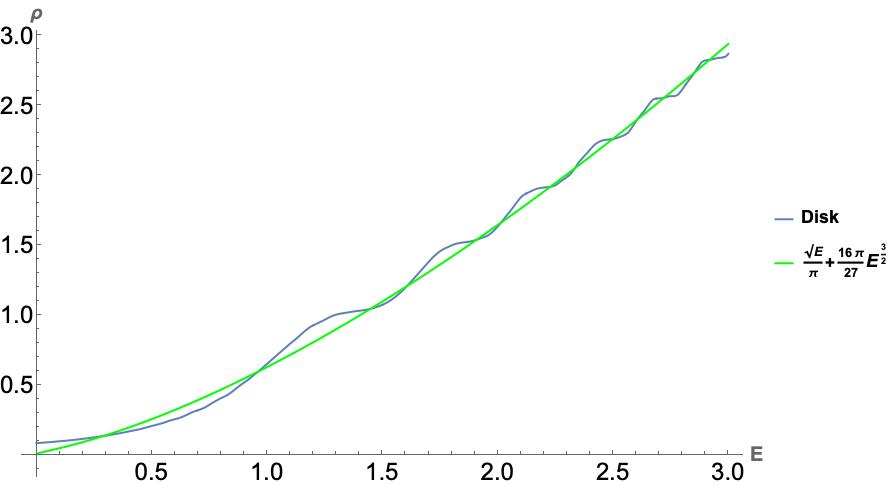}
    \caption{$\rho_0$ vs E curve. We can see that $\rho_0$ perfectly wraps around the expected result in this energy window.}
    \label{fig:disk_density_interp}
\end{figure}

It would be harder than the pure unoriented gravity case to maintain control over the WKB wavefunction in an analytical calculation, so we once again resort to the nonperturbative framework for numerics. Using the numerically determined wavefunctions produces the disk level density shown in fig. \eqref{fig:disk_density_interp}. The expected analytical result, given by
    \begin{equation}
        \rho_0(E) = \frac{\sqrt{E}}{\pi} + \frac{16\pi E^{3/2}}{27},
    \end{equation}
is shown for comparison. 

\begin{figure}[h!]
    \begin{subfigure}[t]{0.49\textwidth}
        \centering
        \includegraphics[width=\textwidth]{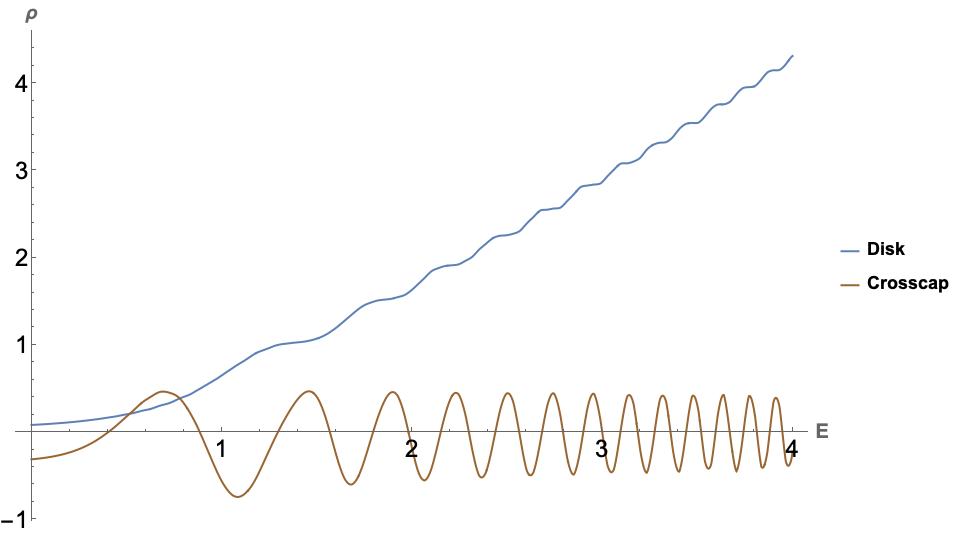}
        \caption{$\rho_0$ and $\rho_{0,1}$ separately}
        \label{fig:crosscap_density_interp}
    \end{subfigure}
    \begin{subfigure}[t]{0.49\textwidth}
        \centering
        \includegraphics[width=\textwidth]{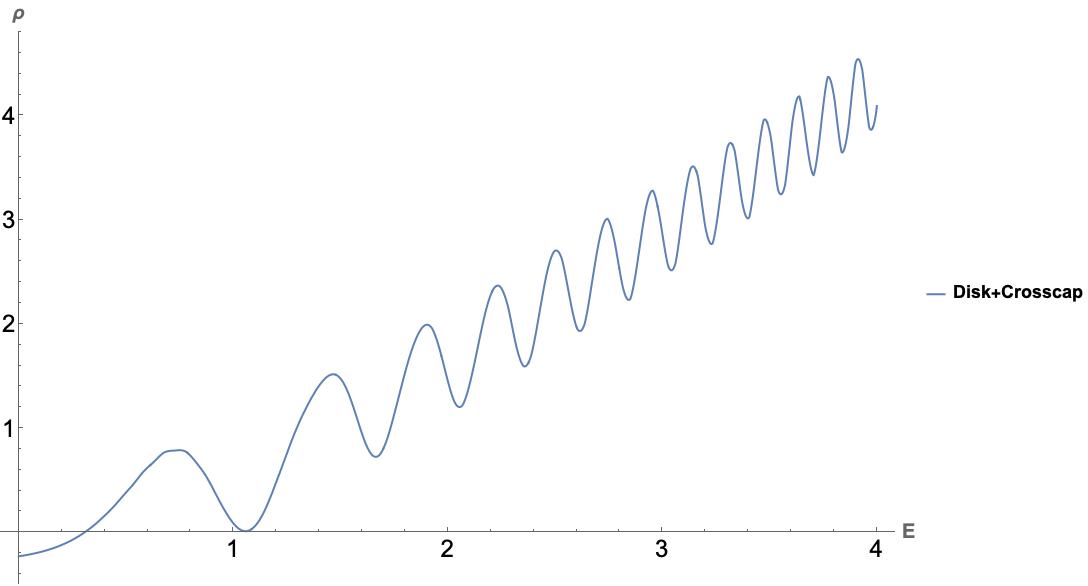}
        \caption{$\rho_0 + \rho_{0,1}$}    
    \end{subfigure}
    \caption{The first unoriented contribution to the eigenvalue density in the interpolated matrix model.}\label{fig:total_density_interp}
\end{figure}

The displayed energy window is smaller in this case than the preceding subsection due to the numerical procedure. The cutoff on the energy in this data is $E_\text{max} \approx 8$. The first perturbative correction can also be calculated using \eqref{crosscapdensity}. The results are displayed in fig. \eqref{fig:total_density_interp}.

\section{Unoriented JT Gravity} \label{scn:JTgravity}

As shown in \cite{Johnson:2020heh,Johnson:2021owr,Johnson:2021tnl,Johnson:2022wsr}, JT gravity is perturbatively defined as a massive interpolation between an infinite number of multicritical models. The full equation of motion obeyed is
    \begin{equation}
        \mathcal{R} \equiv \sum_{k=1}^\infty t_k R_k[u] + x = 0, \label{eqn:stringeq}
    \end{equation}  
where the couplings $t_k$ are determined by the leading density of states of the gravity theory 
    \begin{equation}
        t_k = \frac{\pi^{2k-2}}{2k!(k-1)!}, \label{eqn:orientedcouplings}
    \end{equation}
and $R_k$ is the $k$th Gelfand-Dikii polynomial.

As is clear from the formalism developed above, the function $u$ is seeded from the $\beta = 2$ theory into the $\beta = 1$ theory to determine the functions $g_i$. Further, the communication between the two theories is facilitated by the factorization of the differential operator $P$, the Lax pair of the Schrodinger Hamiltonian. The operator $P$ is a linear combination of $P = \sum_k t_k P_k$ for an interpolation with coupling constants $t_k$. Since the coupling constants are determined by the leading density of states, which is the same in the oriented and unoriented theories (up to normalization), the full operator for unoriented JT gravity will be
    \begin{equation}
        \mathcal{P} = \sum_{k = 1}^\infty \frac{\pi^{2k-2}}{2k!(k-1)!} P_k.
    \end{equation}
We can, in principle, define operators $\mathcal{T},\mathcal{S}$ such that $\mathcal{P} = \mathcal{S}^\dagger \mathcal{T}$ as before, and which map the wavefunctions into the skew wavefunctions. Schematically, the factorization looks like 
    \begin{equation}
        \mathcal{P} = - \prod_{i = 1}^\infty \left( \hbar \partial - \half g_i\right)\hbar \partial \prod_{j = \infty}^1 \left(\hbar \partial + \half g_j\right).
    \end{equation}
The flipped bounds on the right-most product indicate that the functions $g_i$ appear in reverse order in that part
    \begin{equation}
        \mathcal{P} = -\left(\hbar\partial - \half g_1\right)\left(\hbar\partial - \half g_2\right)\cdots \hbar \partial \cdots \left(\hbar\partial + \half g_2\right)\left(\hbar\partial + \half g_1\right).
    \end{equation}
In the same manner as before, we find enough differential equations by comparing coefficients of $\partial$ to determine the $g_i$.

Given the extension of the operator $\mathcal{P}$ to the infinite interpolation that defines JT gravity and its formal factorization in terms of operators $\mathcal{T}$ and $\mathcal{S}$, we predict that (\ref{eqn:density}) remains true for unoriented JT gravity, giving the density
    \begin{equation}
        \rho(E) = \frac{1}{2\hbar} \int_{-\infty}^0dx\Bigg[\mathcal{T}\psi(x,E)\left(\int_0^E \mathcal{S}\psi(x,E')dE'\right) - \mathcal{S}\psi(x,E)\left(\int_E^\infty \mathcal{T} \psi(x,E')dE'\right)\Bigg], \label{eqn:UJTdensity}
    \end{equation}
to all orders in perturbation theory. This formula is both self-consistent and consistent with the infinite interpolation used to define JT gravity in terms of multicritical matrix models.

This theory and orientable JT gravity share the issue of having formally infinite derivatives in the definition. The string equation for $u$ is infinite order, the operator $\mathcal{P}$ is infinite order, and so too are $\mathcal{T}$ and $\mathcal{S}$. The solution is to notice that $t_k$ gets smaller and smaller for higher $k$ and $\lim_{k\rightarrow \infty}t_k=0$. In \cite{Johnson:2020exp} it was noted that a truncation can be made to get reliable numerical results depending on the maximum energy one is interested in probing while computing the eigenvalue density. Practically, one studies the finite interpolation up to some $k_{\text{max}}$ by finding the function $u$ satisfying\footnote{This is only true perturbatively. There are difficulties associated to restricting oneself to perturbation theory, as discussed earlier in the $k = 2$ example. The truncation performed in \cite{Johnson:2020exp} was for the non-perturbative formulation of JT gravity.}
    \begin{equation*}
        \sum_{k = 1}^{k_{\text{max}}} t_k R_k + x = 0,
    \end{equation*}
where $t_k$ are the same coupling constants for JT gravity \eqref{eqn:orientedcouplings}. Then the Schrodinger equation for $u$ can be solved for the wavefunctions.

We believe that, in principle at least, the same can be done in unoriented JT gravity. The algorithm for computing the density is as follows. First, perform the truncated analysis to find $u$ and the wavefunctions $\psi$. Then determine the interpolated operator $\mathcal{P} \approx \sum_{k = 1}^{k_{\text{max}}}P_k$ in terms of $u$, and use that to find the $k_\text{max} - 1$ functions $g_i$. Then build the truncated operators $\mathcal{T}$ and $\mathcal{S}$ out of the $g_i$, and finally use the formula for the density \eqref{eqn:UJTdensity}.

\section{Discussion}

The purpose of this paper was to further explore the study of double scaled $\beta = 1$ Wigner-Dyson models initiated in papers like \cite{Brezin:1990dk, HARRIS1990384}. Recent advancements in two-dimensional quantum gravity have brought to the forefront statistical properties of the theories, with a large part being played by the eigenvalue density. One model, JT gravity, has been of particular importance, and the last several years have seen the development of a cottage industry involving generalizations, deformations, and refinements of that theory. 

Stanford and Witten \cite{Stanford:2019vob} have explored many such extensions of JT gravity from the matrix model perspective, including its definition on unoriented surfaces. Meanwhile, Johnson and others have developed a complementary way of defining JT gravity and its generalizations via a certain combination of simpler matrix models. It is in the spirit of the latter that this work has proceeded.

We have shown that the modern perspective on double scaled matrix models, e.g. a focus on the eigenvalue density, is compatible with the results obtained in the 90s for $\beta = 1$ models, by unifying the more popular approach in the physics community at the time -- orthogonal polynomials -- with the more popular approach in the integrable systems at the time -- skew orthogonal polynomials. 

While it is possible to formally define the objects necessary to compute the eigenvalue density, we found that it is difficult to extract results for specific models. We believe the primary source of difficulty is the fact that the eigenvalue density is non-local in energy space via the presence of the energy integrals. Their mere presence is enough to render the WKB approach to perturbation theory intractable. Despite appealing to the (usually) powerful numerical approach, we found that many efforts to reduce error in one place, e.g. the fidelity of the wavefunctions, resulted in increased error elsewhere, e.g. the energy integrals in $\rho$. Further, the numerics displayed above had $\hbar = 1$, which is not particularly small. There are two logistical problems associated to decreasing $\hbar$: first, the string equation for $u$ is actually harder to solve; second, the wavefunctions oscillate more rapidly, making the diagonalization problem in solving the Schrodinger equation more resource-intensive. 

Despite the numerical difficulties, the leading order contribution to the density in the $k = 2$ model has the desired behavior for small $E$ -- albeit with some fictitious non-perturbative oscillation. The crosscap contribution to the density behaves in accordance with what one would expect of a pertubative correction with a relatively large value of $\hbar$.

As we have pointed out, analysis of the double scaled GOE yields a puzzle: it is missing unoriented contributions. Although it is not a genuine theory of surfaces, one can show that before double scaling there are non-vanishing unoriented contributions to the eigenvalue density when the theory is coupled to quarks\footnote{This is actually a problem in Zee's textbook \cite{zee2010quantum}.}. It would be interesting to explore this further.

Given that the minimal model construction in $\beta = 2$ theories is so intimately connected to the KdV hierarchy, there must be a similar story behind the scenes of the $\beta = 1$ construction above. Two promising avenues to pursue are Drinfeld-Sokolov systems \cite{Drinfeld:1984qv}, due to the fundamental dependence of the double scaled $\beta = 1$ theory on factorizing a Lax operator, and the Pfaff lattice of Adler, et al \cite{adler1999pfaff}, due to its fundamental connection with the $\beta = 1$ matrix integral at finite $N$.

Finally, one should expect that a non-perturbative treatment of the $\beta = 1$ Wigner-Dyson models is possible. The $\beta = 2$ Wigner-Dyson models find as a possible non-perturbative completion the Altand-Zirnbauer (AZ) $(\alpha,\beta) = (1 + 2\Gamma,2)$ models, where the parameter $\Gamma$ can be interpreted as the number of background branes in the theory \cite{Dalley:1991vr, Johnson:2019eik, Johnson:2020exp, Johnson:2021tnl} . The logical prediction is that the AZ $(\alpha,\beta) = (1 + \gamma,1)$ models, for constant $\gamma$, will provide a non-perturbative completion of the models considered here. Demonstrating the connection between the $\beta = 2$ and $(\alpha,\beta) = (1+2\Gamma,2)$ models requires the use of the KdV integrable structure of the theory, so it is possible that a non-perturbative completion of the $\beta = 1$ models will require a more careful treatment of their integrable structure as well. It would be interesting if such a non-perturbative completion of the unoriented theory could be used to study O-planes in two dimensions.

\acknowledgments

We are grateful to Clifford V. Johnson for many helpful conversations and his guidance in the preperation of this manuscript. Wasif Ahmed thanks Sushmit Hossain for helping initially with MATLAB. This work is supported in part by the US Department of Energy under grant DE-SC0011687.


\appendix

\section{The GOE}

\subsection{Saddle point analysis}

For any $\beta$ and matrix potential $V$, the matrix integral of the Dyson $\beta$-ensemble is
    \begin{equation}
        \tilde{Z} = \int \prod_i d\lambda_i \exp\left\{-N\sum_i V(\lambda_i) + \frac{\beta}{2}\sum_{i \neq j} \log|\lambda_i - \lambda_j|\right\}.
    \end{equation}
Introducing the continuous spectral density $\tilde{\rho}_0$ in the usual fashion and passing to the large-$N$ limit, the exponential in the integrand can be written
    \begin{equation}
        -N^2\left[ \int \tilde{\rho}(\lambda)V(\lambda)d\lambda - \frac{\beta}{2}\iint \tilde{\rho}(\lambda)\tilde{\rho}(\mu)\log|\lambda - \mu|d\lambda d\mu \right].
    \end{equation}
The large-$N$ saddle point solution results from the condition
    \begin{equation}
        V'(\lambda) = \beta P \int_{-a}^a \frac{\tilde{\rho}(\mu)}{\lambda - \mu} d\mu.
    \end{equation}

Next, introduce the analytic function 
    \begin{equation}
        F(\lambda) = \int \frac{\tilde{\rho}(\mu)}{\lambda - \mu},
    \end{equation}
which, via application of the Sokhotski–Plemelj formula, has the discontinuity
    \begin{equation}
        F(\lambda \pm i \epsilon) = \frac{V'(\lambda)}{\beta} \mp i \pi \tilde{\rho}(\lambda).
    \end{equation}
A solution for $\tilde{\rho}$ is found by using the ansatz
    \begin{equation}
        F(\lambda) = \frac{V'(\lambda)}{\beta} - Q(\lambda){\pi}\sqrt{\lambda^2 - a^2},
    \end{equation}
where the polynomial $Q$ and the endpoints $\pm a$ are determined by requiring that $F \to 1/\lambda$ as $|\lambda| \to \infty$. The density is then given by
    \begin{equation}
        \tilde{\rho}(\lambda) = \frac{Q(\lambda)}{\pi} \sqrt{a^2 - \lambda^2}.
    \end{equation}

For the Gaussian case $V(\lambda) = \alpha \lambda^2$ one finds
    \begin{equation}
        Q(\lambda) = \frac{2\alpha}{\beta} \quad \& \quad a^2 = \frac{\beta}{\alpha},
    \end{equation}
so that the density is 
    \begin{equation}
        \tilde{\rho}(\lambda) = \frac{2\alpha}{\pi \beta}\sqrt{\frac{\beta}{\alpha} - \lambda^2} = \frac{2}{\pi a^2}\sqrt{a^2 - \lambda^2}. \label{eqn:generaldensity}
    \end{equation}

Given the canonical normalization $\alpha = \half$, we have the following results for $\beta = 2,1$ (respectively)
    \begin{equation}
        \begin{aligned}
            \tilde{\rho}(\lambda) &= \frac{\sqrt{4 - \lambda^2}}{2\pi}, \\
            \tilde{\rho}(\lambda) &= \frac{\sqrt{2 - \lambda^2}}{\pi}.
        \end{aligned}
    \end{equation}
Upon performing double scaling with the relation $\lambda = \lambda_c^{(\beta)} - \delta^2 E$, with $\lambda_c^{(2)} = 2$ and $\lambda_c^{(1)} = \sqrt{2}$, we have
    \begin{equation}
        \begin{aligned}
            \rho(E) &= \frac{\sqrt{E}}{\pi\hbar}, \\
            \rho(E) &= \frac{2^{3/4}\sqrt{E}}{\pi\hbar}, \label{eqn:scaleddensity}
        \end{aligned}
    \end{equation}
for $\beta = 2,1$ respectively\footnote{Some compensating factors of $N$ need to be included to extract the scaling part of the density. These factors of $N$ combine with the left over power of $\delta$ to form the renormalized parameter $\hbar$.}.

Notice that the $\beta = 1$ density has the same functional form as the $\beta = 2$ density. Indeed, if we plug $\beta = 2$ and $\alpha = 1$ into eq. (\ref{eqn:generaldensity}) we obtain the lower result in eq. (\ref{eqn:scaleddensity}). This is neither a surprise nor an accident. Of course we expect both ensembles to have the same leading dependence on $E$ (i.e. the Wigner semi-circle), but the exact relationship between the normalizations can be anticipated by looking at orthogonal polynomials.

\subsection{The crosscap resolvent and matrix model recursion}

For general $\beta$, the recursion relation determining the resolvent at genus $g = 0,\half,1,\cdots,$ and number of boundaries $n$ is \cite{Stanford:2019vob}
    \begin{equation}
        \sqrt{\sigma(\lambda)} R_{g,n}(\lambda_1,\cdots,\lambda_n) = \frac{1}{2\pi i} \int_{\mathcal{C}} \frac{d\lambda}{\lambda - \lambda_1} F_{g,n}(\lambda,I) \frac{\sqrt{\sigma(\lambda)}}{2y(\lambda)},
    \end{equation}
where $\sigma(\lambda) = \lambda^2 - a^2$, $\mathcal{C}$ is a contour surrounding the cut $[-a,a]$, $y(\lambda)$ is the spectral curve, $I$ denotes the set $\{\lambda_2,\cdots,\lambda_n\}$, and
    \begin{equation}
        \begin{aligned}
            F_{g,n}(\lambda,I) &= \left(1 - \frac{2}{\beta} \right) \partial_\lambda R_{g - \half,n}(\lambda,I) + R_{g-1,n+1}(\lambda,\lambda,I) \\
            &\quad + \sum_{\text{stable}}R_{h,1+|J|}(\lambda,J)R_{h',1+|J'|}(\lambda,J') \\
            &\quad + 2\sum_{k = 1}^n\left(R_{0,2}(\lambda,\lambda_k) + \frac{1}{\beta} \frac{1}{(\lambda - \lambda_k)^2}\right)R_{g,|I|-1}(\lambda,I\setminus \lambda_k).
        \end{aligned}
    \end{equation}
The subscript `stable' on the sum means we ignore terms where one of the factors is $R_{0,1}(\lambda)$ or $R_{0,2}(\lambda,\lambda_k)$.

The only term relevant to the crosscap (with one boundary) is the first one. We have
    \begin{equation}
        F_{\half,1}(\lambda) = -\partial_\lambda y(\lambda).
    \end{equation}
The spectral curve is proportional to the leading eigenvalue density 
    \begin{equation}
        y(\lambda) = -i \pi \tilde{\rho}(\lambda).
    \end{equation}
For the GOE, the spectral curve $y$ is also proportional to $\sqrt{\sigma}$, as can be seen in the saddle point analysis above. Before proceeding explicitly, we perform double scaling $\lambda \to -a + E$, after which the spectral curve is $y(E) \propto \sqrt{-E}$ which is pure imaginary on the positive real axis. In terms of the double scaled spectral curve, the factor $F$ in the integrand is
    \begin{equation}
        F_{\half,1}(E) \propto \frac{1}{\sqrt{-E}}.
    \end{equation}
The contour $\mathcal{C}$ is formally a loop around the positive $E$-axis where the branch cut is. Depending on whether or not the integrand is single-valued, we will either be able to take the residue at the origin or use a standard keyhole contour. As an integral over a real variable, the domain of integration is the nonnegative real line, $\mathbb{R}^+$.

The double scaled resolvent is given by
    \begin{equation}
        \sqrt{-E_1}R_{\half,1}(E_1) \propto \int_{C}\frac{dE}{E - E_1} \frac{1}{\sqrt{-E}}.
    \end{equation}
The integrand has a simple pole at $E = E_1$ and a branch cut
from $E = 0$ to $E = \infty$. Since it is not single valued, the integral cannot be written as a residue at the origin.

One can derive a general formula for the integral
    \begin{equation}
        I(f;a) = \int_0^\infty dx f(x) x^a,
    \end{equation}
where $a > -1$ is a non-integer real number and $f$ is a function with simple poles on the positive real axis that satisfies
    \begin{equation}
        |f(R)| \leq \frac{M}{R^p},
    \end{equation}
with $p > a + 1$ and $R \gg 1$\footnote{This is enough (along with $a > -1$) to ensure that the integral vanishes on the circles around the origin and infinity.} (see for example \cite{hayek2000advanced}). Do this by passing to the complex plane with the complex coordinate $E$ and considering a generalization of the keyhole contour. Beginning with the standard keyhole contour, we make indentations in the parts parallel to the positive real axis at the poles of the function $f$. The result is
    \begin{equation}
        I(f;a) = -\frac{\pi e^{-i\pi a}}{\sin(\pi a)}\sum_j r_j + \pi \text{cotan}(\pi a) \sum_j r^\ast_j,
    \end{equation}
where $r_j$ is the residue of $f(x)x^a$ at the $j^\text{th}$ pole in $\mathbb{C}\setminus \mathbb{R}^+$ and $r^\ast_j$ is the residue at the $j^\text{th}$ pole on $\mathbb{R}^+$ evaluated above the branch cut.

The crosscap resolvent is clearly proportional to the integral $I\left(f;-\half \right)$ with the function $f(E) = 1/(E-E_1)$. For $R\gg 1$ and $R \gg E_1$ we have $|f(R)| \sim 1/R$, and $1 > -\half + 1$. Therefore we can apply the above formula. Since we take $E_1 \in \mathbb{R}^+$, there are no residues in the rest of the complex plane. But, since cotan$(-\pi/2) = 0$, the second term in $I$ is identically 0. Therefore we conclude that 
    \begin{equation}
        R_{\half,1}(E) = 0.
    \end{equation}

To see this another way, introduce the uniformizing coordinate $z$ via $z^2 = -E$. Then
    \begin{equation}
        z_1R_{\half,1}(z_1) \propto \int_C \frac{dz}{z^2 - z_1^2}.
    \end{equation}
The factor of $1/z$ in the integrand is cancelled by the Jacobian of the change of variables. If after the mapping the contour $\mathcal{C}$ contains both $\pm z_1$, then we find
    \begin{equation}
        R_{\half,1}(z) = 0,
    \end{equation}
because the residues contribute with opposite signs. If the contour contains neither of $\pm z_1$, then the resolvent is identically 0. In both cases, the crosscap density is 0.


\section{Pseudo-differential Operators} \label{apdx:PDO}

\subsection{Fractional Powers}

The Schrodinger Hamiltonian $\mathcal{H}$ is a second-order differential operator, and so there is a natural way raise it to a half-integer power. Begin by setting
	\begin{equation}
		\mathcal{H}^\half = i\hbar\partial + \sum_{n = 0}^\infty q_{-n} \hbar^{-n} \partial^{-n},
	\end{equation}
where a negative power of $\partial$ represents an antiderivative, and $q_{-n}$ will be a polynomial in $u$ and its derivatives. The generalization of the Leibniz rule to antiderivatives is given by
	\begin{equation}
		\partial^{-n} f = \sum_{i = 0}^\infty \binom{-n}{i} f^{(i)} \partial^{-n-i},
	\end{equation}
where $f$ is a function of $x$. Then by utilizing this generalized Leibniz rule, the coefficient functions $q_{-n}$ are determined term-by-term by imposing $\left(\mathcal{H}^\half \right)^2 = \mathcal{H}$. The first several are
    \begin{equation}
    \begin{aligned}
        q_0 &= 0, \mspace{100mu} q_{-1} = -\frac{i}{2}u,\\
        q_{-2}&=\frac{i\hbar}{4}u', \mspace{75mu} q_{-3}= -\frac{i}{8}\left(\hbar^2 u'' + u^2\right).
    \end{aligned}
    \end{equation}  

\subsection{The Momentum Operator}

Once we have $\mathcal{H}^\half$, the higher half-integer powers are obtained by $\mathcal{H}^{k - \half} = \mathcal{H}^{k-1} \mathcal{H}^\half$. 

The highest order derivative term will be obtained from 
    \begin{equation}
(-\hbar^2\partial^2)^{k-1} \times i\hbar \partial = i(-1)^{k-1}\hbar^{2k-1}\partial^{2k-1}.
    \end{equation}
The next non-vanishing term will be proportional to $\partial^{2k-3}$. We can think of there being two contributions to this. The first is 
    \begin{equation}
        (-\hbar^2\partial^2)^{k-1} \times \hbar^{-1}q_{-1}\partial^{-1},
    \end{equation}
where all of the derivatives are commuted past $q_{-1}$ to leave the desired power of $\partial$. The coefficient from this term is $(-1)^{k}\frac{i\hbar^{2k-3}}{2}u$. The second contribution comes from the next highest order cross-term in $\mathcal{H}^{k-1}$, which is obtained by replacing one factor of $-\hbar^2 \partial^2$ with $u$, and commuting the derivatives to the right
    \begin{equation}
        u(-\hbar^2 \partial^2)^{k-2} \times i\hbar \partial.
    \end{equation}
There are $k-1$ ways to do this, giving the coefficient $(-1)^{k}(k-1)i\hbar^{2k-3} u$. Therefore the operator $P_k$ is
    \begin{equation}
        -i P_k = (-1)^{k-1} \hbar^{2k-1}\partial^{2k-1} + (-1)^k \left(k - \half\right)u \hbar^{2k-3}\partial^{2k-3} + \cdots.
    \end{equation}  
This justifies the coefficient on the left hand side of \eqref{eqn:geqn1}.

Based on the discussion above, we may be interested in linear combinations of $P_i$ for different values of $i$. For instance, consider the superposition
    \begin{equation}
        P = \sum_{j = 1}^k t_j P_j,
    \end{equation}
for some $k \geq 2$. The highest order derivative in $P_j$ is strictly increasing with $j$, so the maximum power of $\partial$ appearing in $P$ will be $\partial^{2k-1}$, and the only term in the superposition that can contribute to that term is $P_k$. The next power of $\partial$ appearing in $P$ is $\partial^{2k-3}$, whose coefficient will receive contributions from $P_k$ and $P_{k-1}$. Thus
    \begin{equation}
        -iP = (-1)^{k-1} t_k \hbar^{2k-1}\partial^{2k-1} + (-1)^k \Bigg[ \left(k - \half\right) t_k u + t_{k-1} \Bigg]\hbar^{2k-3}\partial^{2k-3} + \cdots.
    \end{equation}
Note that no other terms in the superposition can contribute at these orders.

\subsection{Factorization and New String Equations}

Consider the factorization of $P_k$ in eq. \eqref{eqn:factorization}, displayed again here for convenience
    \begin{equation}
    P_k =(-1)^{k-1}it_k\prod_{i = 1}^{k-1}\left(\hbar\partial - \half g_i\right)\hbar\partial \prod_{j = 1}^{k-1}\left(\hbar\partial + \half g_{k-j}\right). \label{eqn:factorization2}
\end{equation}
The first condition on the functions $g_j$ will be found by comparing the coefficients of $\partial^{2k-3}$ on both sides. In the preceding subsection we determined this coefficient in $P$ for an arbitrary superposition.  

There are two types of term that show up in the coefficient of $\partial^{2k-3}$ on the right hand side of \eqref{eqn:factorization2}. The first comes from commuting all derivatives past functions. Since none of the functions will be differentiated, the mixed terms involving $g_ig_j$ with $i \neq j$ will cancel by symmetry. This leaves 
    \begin{equation}
        \hbar^{2k-3} \left(\sum_{i=1}^{k-1} g_i^2\right)\partial^{2k-3}.
    \end{equation}
The other type of term comes from letting one of the derivatives in the factorization act on one of the $g_i$. The possible outcomes can be visualized by breaking the factorization up into the operators $T^\dagger$ and $S$ as in eq. \eqref{eqn:TandSdefs}, corresponding to lumping the terms of the form $\prod(\partial - g)$ into $T^\dagger$ and the ones of the form $\partial \prod (\partial + g)$ into $S$. The function $g_j$ can be differentiated once $k - j$ times in $T^\dagger$ and $j+k$ times in $S$, since the terms in $S$ can also be differentiated by the terms in $T^\dagger$. Taking into account the minus signs and factors of 2, this gives 
    \begin{equation}
        -\hbar^{2k-3} \left(\sum_{j=1}^{k-1} jg_j'\right)\partial^{2k-3}.
    \end{equation}
Hence the factorization is
    \begin{equation}
    \begin{aligned}
        (-1)^{k-1}it_k\prod_{i = 1}^{k-1}\left(\hbar\partial - \half g_i\right)\hbar\partial & \prod_{j = 1}^{k-1}\left(\hbar\partial + \half g_{k-j}\right) = (-1)^{k-1}it_k\hbar^{2k-1}\partial^{2k-1} \\
        &+ (-1)^{k-1}i t_k\left( \sum_{j = 1}^{k-1}-jg_j' + \frac{1}{4}g_j^2 \right)\hbar^{2k-3}\partial^{2k-3} + \cdots.
    \end{aligned}
    \end{equation}

\bibliographystyle{JHEP}
\bibliography{biblio.bib}

\end{document}